\documentclass{emulateapj}

\usepackage{graphicx}
\usepackage{epstopdf}
\usepackage{amsmath}
\usepackage{amssymb}
\usepackage{bm}

\def\gs{\mathrel{\raise0.35ex\hbox{$\scriptstyle >$}\kern-0.6em
\lower0.40ex\hbox{{$\scriptstyle \sim$}}}}
\def\ls{\mathrel{\raise0.35ex\hbox{$\scriptstyle <$}\kern-0.6em
\lower0.40ex\hbox{{$\scriptstyle \sim$}}}}

\shorttitle{Geometric and Dynamical Models of Reverberation Mapping Data}
\shortauthors{Pancoast et al.}

\begin{document}

\title{Geometric and Dynamical Models \\
 of Reverberation Mapping Data}

\author{Anna Pancoast\altaffilmark{1}, Brendon J. Brewer\altaffilmark{1}, Tommaso Treu\altaffilmark{1,2}}
\email{pancoast@physics.ucsb.edu}

\altaffiltext{1}{Department of Physics, University of California, Santa Barbara, CA, 93106-9530, USA}
\altaffiltext{2}{Packard Research Fellow}

\begin{abstract}
We present a general method to analyze reverberation (or echo) mapping data, that simultaneously provides estimates for the black hole mass and for the geometry and dynamics of the broad line region (BLR) in active galactic nuclei (AGN).  While previous methods yield a typical scale size of the broad line region or a reconstruction of the transfer function, our method directly infers the spatial and velocity distribution of the BLR from the data, from which a transfer function can be easily derived. Previous echo mapping analysis requires an independent estimate of a scaling factor known as the virial coefficient to infer the mass of the black hole, but this is not needed in our more direct approach. We use the formalism of Bayesian probability theory and implement a Markov Chain Monte Carlo algorithm to obtain estimates and uncertainties for the parameters of our BLR models. Fitting of models to the data requires knowledge of the continuum flux at all times, not just the measured times. We use Gaussian Processes to interpolate and extrapolate the continuum light curve data in a fully consistent probabilistic manner, taking the associated errors into account. We illustrate our method using simple models of BLR geometry and dynamics and show that we can recover the parameter values of our test systems with realistic uncertainties that depend upon the variability of the AGN and the quality of the reverberation mapping observing campaign.  With a geometry model we can recover the mean radius of the BLR to within $\sim0.1$\,dex random uncertainty for simulated data with an integrated line flux uncertainty of $1.5$\%, while with a dynamical model we can recover the black hole mass and the mean radius to within $\sim0.05$\,dex random uncertainty, for simulated data with a line profile average signal to noise ratio of 4 per spectral pixel.   These uncertainties do not include modeling errors, which are likely to be present in the analysis of real data, and should therefore be considered as lower limits to the accuracy of the method. 

\end{abstract}

\keywords{galaxies:active --- methods: data analysis --- methods: statistical}

\section{Introduction}
The energy emitted by active galactic nuclei (AGN) is argued to be the
result of matter accreting onto supermassive black holes at the center
of galaxies \citep{lynden71}.  However, the details of the geometry
and kinematics of the region around the accretion disk are not well
understood .  In the standard model of AGN, the region around the
accretion disk is called the broad line region (BLR) due to the broad
emission lines from rapidly moving clouds of material near the black
hole \citep{antonucci93, urry95}.  Models for the BLR attempt to
explain the many categories of AGN by the observer's viewing angle and
the covering fraction of inflowing or outflowing material \citep[see
e.g.][]{murray95, elvis00}, since the degree to which the BLR geometry
and kinematics vary between individual systems is also unknown.

The mass of the central black hole is a fundamental parameter in galaxy evolution, as suggested by a relation between black hole mass and stellar velocity dispersion of the host galaxy bulge, the $M_{BH} - \sigma_{\star}$ relation \citep[see e.g.][]{bennert11}.  
While the black hole masses of very nearby galaxies can be measured by observing the orbits of stars or gas, a different approach is needed for more distant galaxies because the gravitational sphere of influence of the black hole cannot be spatially resolved \citep[e.g.][]{ferrarese05}. In active galaxies, the small size of the BLR, estimated to be around $\sim 10^{14}-10^{16}$m \citep{wandel99, kaspi00, bentz06}, inhibits direct imaging of the accretion disk and orbiting BLR clouds.

Reverberation mapping provides a method to determine the black hole mass, along with the geometry and the kinematics of the BLR \citep{blandford82, peterson93, peterson04}.  Without relying on spatially resolving the gravitational sphere of influence of the black hole, reverberation mapping provides a powerful tool for studying black holes over a range of redshifts and black hole masses \citep[e.g.][]{peterson04, woo07, bentz09, denney09}.  The method relies on the large time-variability of AGN luminosity, spanning timescales of days to years \citep[e.g.][]{webb00}.  Reverberation mapping data consists of a timeseries of frequent measurements of the intensity of the continuum and broad line emission.  The line emission strength is assumed to be proportional to the continuum emission strength, but with a time lag due to the light travel time from the central ionizing source to the BLR material.  The lag time between line and continuum flux contains information about the size of the BLR, while the shape of the spectral line encodes the velocity information.  An estimate of the black hole mass can be calculated assuming the BLR clouds orbit in the Keplerian potential of the black hole with velocities determined by the width of the spectral line and at a radius given by the average lag between the line and continuum fluxes.

A weakness of this traditional reverberation mapping method is that the relation between velocity and position observables of the clouds and the black hole mass depends on an unknown dimensionless proportionality constant that depends on the geometry and kinematics of the BLR.  In practice, this so-called ``virial'' factor is estimated based upon some external criteria, such as the average factor that makes the $M_{BH}-\sigma_{\star}$ relation consistent between different black hole mass estimators and between samples of active and inactive galaxies \citep{onken04, collin06, woo10, greene10, graham10}.  Ideally, we would like to infer directly the morphology and kinematics of the BLR, and the black hole mass, including its uncertainty \citep[for a discussion of potential systematic errors in reverberation mapping see][]{krolik01}.  The models for the structure and kinematics may include a net inflow or outflow of BLR clouds, among other physically motivated models \citep[e.g.][]{murray95, marconi08}. Developing such a method is the goal of this paper.

The data required for reverberation mapping encode the geometry and kinematic information to some degree, depending upon the quality, in the form of the transfer function (or response function) that maps the continuum emission onto the line emission.  The average lag used to estimate black hole mass is the first moment of the transfer function. Previous analysis involved estimating the transfer function and then interpreting the transfer function in relation to a model of the BLR \citep[see][]{krolik95, done96, horne03, bentz10}.  This is necessary because the transfer function is a function of time lag, not position within the BLR, so its interpretation requires a physical model.  Estimating the transfer function requires inverting a linear integral equation, and while the method of \citet{krolik95} and \citet{done96} uses regularized linear inversion, thus allowing for uncertainty estimation, other inversion methods such as ``maximum entropy'' do not allow for straightforward uncertainty estimation or model selection \citep[e.g.][]{horne03, bentz10}.  

Our method of analyzing reverberation mapping data simplifies the process of obtaining a transfer function and then interpreting the result using different models. We compare reverberation mapping data directly with models of the broad line region, obtaining uncertainty estimates as well as allowing for model selection.  Once we have found models and model parameters that fit the data, we can easily compute the transfer function and average time-lag.  Our goal is to constrain the geometry and kinematics of the BLR and provide an internally consistent factor for the black hole mass.  We note that the traditionally determined average time-lag is exactly equivalent to a model where the BLR is a face-on ring of a given radius (response = $\delta$-function) or a spherical shell (response = step function). This implicit assumption drives the inference on the average lag and its result, as we will show in this paper.

An important part of our method for directly modeling re-
verberation mapping data requires that we predict the AGN
continuum light curve between the observations. Recent work
has found that AGN continuum light curves are well-modeled
by a damped random walk: \citet{2009ApJ...698..895K} used a continuous
time stochastic process; \citet{2010ApJ...708..927K} used the formalism of \citet{1992ApJ...385..404P} and \citet{1992ApJ...398..169R,1994comp.gas..5004R}.
This model for AGN variability was applied to ∼900 AGNs
by \citet{2010ApJ...721.1014M} in order to correlate variability with
other parameters of AGNs.  \citet{zu10} were then able to use
this model for AGN variability to improve the standard analysis
of reverberation mapping data, including a better understanding of the uncertainties involved. They model the continuum
light curve using Gaussian Processes to recover the transfer
function, assumed to be a top-hat. As with \citet{2010ApJ...708..927K}, they use an exponential covariance matrix to relate the
continuum flux at different points in the time series. We also
use Gaussian Processes to model the continuum light curve, as
well as a slightly more general exponential covariance matrix.
Our method improves upon the approach of \citet{zu10} by
modeling the reverberation mapping data directly in terms of
a geometric and dynamical model, rather than recovering the
transfer function.

\citet{bottorff97} have also modeled reverberation mapping directly in an attempt to understand the BLR dynamics in the well-studied AGN NGC 5548.  They expand upon the hydromagnetically driven outflow model of \citet{emmering92} and use one set of parameter values to compare their model with NGC 5548.  While the specific models presented here are clearly not as sophisticated from a physical point of view, our method improves upon that approach by finding the best fit parameter values of our simple models and believable estimates of their uncertainties.

We consider two types of reverberation mapping data sets: velocity-unresolved, where there is a time series of the continuum flux and a time series of the integrated line flux, and velocity-resolved, where the data consist of a continuum flux time series, and a series of entire line spectra as a function of time.

The paper is organized as follows.  In \S\,\ref{sect_pic} we define and describe the physical problem.  In \S\,\ref{sect_method} we outline our methods in the formalism of Bayesian probability theory and describe the algorithms we use to compare reverberation mapping data to mock data created from a model of the BLR.  In \S\,\ref{sect_tests} we test our method using simple models of the BLR and show that we are able to recover the parameter values of our test systems.  Finally, in \S\,\ref{sect_concl}, we summarize our conclusions.  Flux units throughout the paper are arbitrary, but computed consistently within our method.

\begin{figure*}
\begin{center}
\includegraphics[scale=0.8]{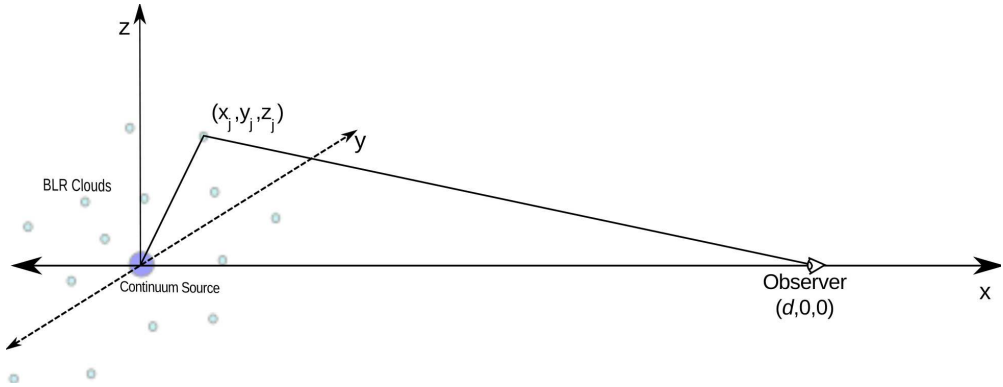}
\caption{BLR clouds around the central ionizing source (central engine). The extra path length the light must travel from the central engine to the BLR cloud and then to the observer is the cause of the delayed response of the line flux.\label{diagram}}
\end{center}
\end{figure*}

\begin{figure*}
\begin{center}
\includegraphics[scale=0.75]{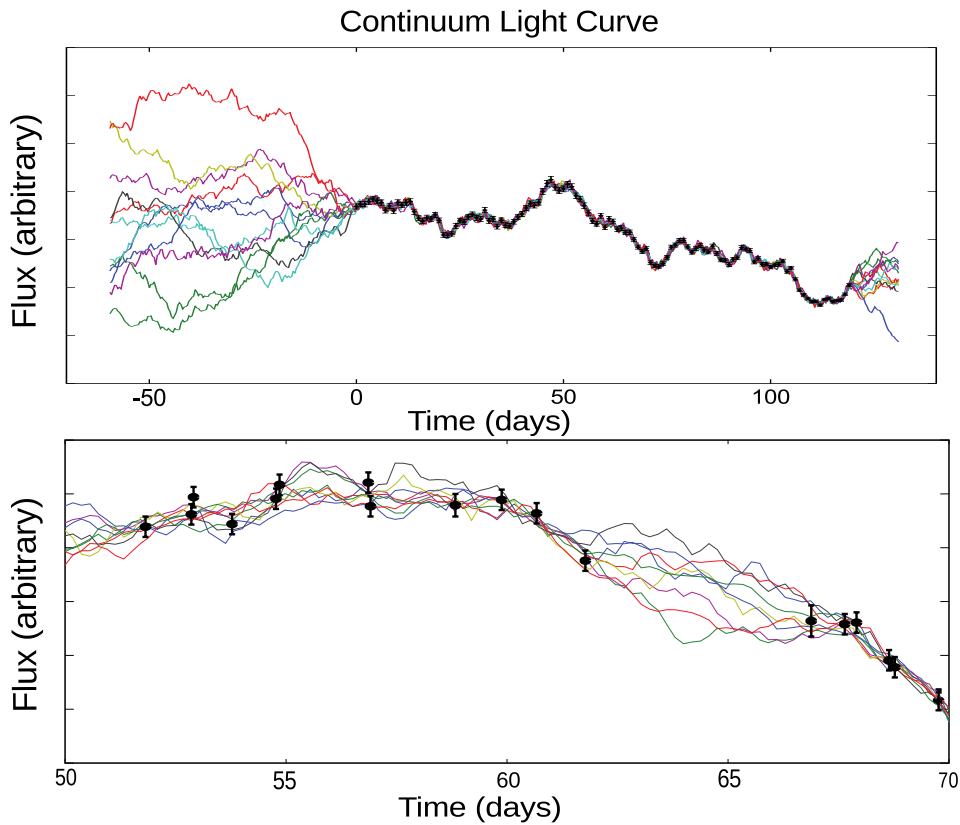}
\end{center}
\caption{Simulated continuum emission datapoints with examples of the
continuum interpolated using gaussian processes. The dispersion
of the lines represents the uncertainty of the recovered light
curve. As expected the uncertanty is greatest where there are no data
points. The top panel shows the simulated data used throughout this
paper, whereas the bottom panel shows an example with gaps in the
data. Our procedure takes into account the amount of information
available and therefore the recovered light curve suffers from a
larger uncertainty during the gaps.
\label{continuum}}
\end{figure*}

\section{The Physical Picture}
\label{sect_pic}
Throughout this paper, we assume a simple model for the BLR, described as follows. The AGN is defined to be at $(0,0,0)$, and the observer is at $(d, 0, 0)$. We model the distribution of BLR gas by defining the gas density profile $\rho(x,y,z)$, assumed to be normalized such that
\begin{equation}
\int_V \rho(x, y, z) \, dV = 1
\end{equation}
where $dV = dx \, dy \, dz$ and $V$ is all of space.  We assume that the gas absorbs the ionizing radiation, but is not self-shielding, so that gas at larger radii is still illuminated.  It should be noted that our approach is fully general and can support more complex models of the optical properties of the BLR, as well as its geometry and dynamics.

\subsection{Velocity-Unresolved Reverberation Mapping}
If the continuum flux varies with time according to $f_{\rm cont}(t)$, then the total line flux as a function of time is given by
\begin{equation}\label{transfer}
f_{\rm line}(t) = A \int_V \ f_{\rm cont}(t - l(x, y, z)) \rho(x, y, z) \, dV
\end{equation}
where $l(x, y, z)$ is the {\it lag}, or time delay, associated with BLR gas at position $(x, y, z)$, and $A$ is a response coefficient. The lag $l$ for each position is simply the excess light travel time from taking a path starting at $(0,0,0)$ that travels to some gas at $(x, y, z)$, where the light is absorbed and reemitted as line emission, and that finally travels to the observer, relative to a direct path straight from the AGN to the observer: 
\begin{eqnarray}
l(x,y,z) = \left(\sqrt{x^2 + y^2 + z^2}\right. \\
+ \left.\sqrt{(x - d)^2 + y^2 + z^2} - d\right)/c
\end{eqnarray}
For any case of interest, $d \gg \sqrt{x^2 + y^2 + z^2}$, and therefore this is well approximated by:
\begin{equation}
l(x,y,z) \approx \left(\sqrt{x^2 + y^2 + z^2} - x\right)/c
\end{equation}
which is the formula adopted throughout this paper. See Figure~\ref{diagram} for an illustration of this model.

Note that Equation~\ref{transfer} is a special case of the general equation
\begin{equation}\label{transfer2}
f_{\rm line}(t) = A \int \Psi(\tau) f_{\rm cont}(t - \tau) \, d\tau
\end{equation}
where $\Psi(t)$ is the so-called {\it transfer function}, which gives the response of the line flux to a delta-function pulse in the continuum flux\footnote{For readers more familiar with image analysis, the transfer function is analogous to a PSF.}. Thus, for any particular system, if we can infer the density of BLR clouds throughout space, we can automatically deduce the corresponding transfer function:
\begin{equation}
\Psi(\tau) = \int_V  \delta\left(\tau - l(x, y, z)\right)\rho(x, y, z) \, dV
\end{equation}
The meaning of this equation is that each location in space contributes to the transfer function at the value of the location's lag, with the size of the contribution being proportional to the amount of gas at that location.

\subsection{Velocity-Resolved Reverberation Mapping}
Now suppose that the BLR gas is in motion, such that the system can be described by a time-invariant distribution function $g$ defined over the phase space of a single particle:
\begin{equation}
g(x, y, z, v_x, v_y, v_z) = \rho(x, y, z)g(v_x, v_y, v_z | x, y, z)
\end{equation}
The motion of the gas along the line of sight is assumed to affect the wavelength of reemitted light, but its distribution function is assumed to be time invariant and therefore does not vary during the observing campaign. Then the emission line profile at time $t$ will be a function of the line of sight velocity, $v_{\rm los}$: 
\begin{eqnarray}
f_{\rm line}(v_{\rm los},t) = A \int_{v_y, v_z} \int_V \ f_{\rm cont}(t - l(x, y, z)) \\ \times g(x, y, z, v_x, v_y, v_z) \, dx \, dy \, dz \, dv_y \, dv_z
\end{eqnarray}
where $v_{\rm los}$ is in the $x$ direction.  This is the velocity-resolved equivalent of Equation~\ref{transfer}.

\section{Method}
\label{sect_method}
Our method for constraining the geometry and kinematics of the BLR is an application of Bayesian Inference \citep{sivia06}. In general, to infer parameters $\theta$ from data $D$, we begin by assigning a prior probability distribution $p(\theta)$ describing our initial uncertainty about the parameters. Sampling distributions $p(D|\theta)$ are also assigned to describe our uncertainty about how the data are related to the parameters. Once specific data $D=D^*$ are obtained, our updated state of knowledge about the parameters is described by the posterior distribution, given by Bayes' rule:
\begin{equation}
\label{eqn_posterior}
p(\theta | D=D^*, I) \propto p(\theta|I) p(D|\theta, I)|_{D=D^*}
\end{equation}
Here $I$ is any background information we have about the problem.  In complex problems, where $\theta$ consists of a large number of parameters, Monte Carlo methods are used to produce random samples from the posterior distribution for $\theta$. Methods such as Nested Sampling \citep{dnest} can also provide the normalization constant for the posterior, known as the evidence, which is the key quantity for comparing the entire model with an alternative \citep{sivia06}.

In our method, the parameters $\theta$ to be inferred are those describing the spatial profile of the BLR gas, and the continuous continuum flux, $f_{\rm cont}(t)$. Since it is impossible to represent a continuum in a computer, we instead infer $f_{\rm cont}(t)$ evaluated at 500 time points, covering a time interval larger than the continuum data.  The continuum modeling technique is described in detail in the next section.

Throughout this paper, both the continuum flux and line flux timeseries are considered part of the dataset $D$:
\begin{equation}
D = \left\{ \mathbf{y}_{\rm line}, \mathbf{y}_{\rm continuum} \right\}
\end{equation}
The prior information consists of the times at which the line flux and continuum flux are measured, $\mathbf{t}$ and the error bars on the line flux and continuum flux measurements, $\bm{\sigma}$:
\begin{equation}
I = \left\{ \left(\mathbf{t}, \bm{\sigma}\right)_{\rm line}, \left(\mathbf{t}, \bm{\sigma}\right)_{\rm continuum}\right\}
\end{equation}

The likelihood function is chosen to be Gaussian, centered around the model-predicted line flux timeseries:
\begin{equation}
p(D|\theta) = \prod_{i=1}^n \frac{\exp\left[-\frac{1}{2}\left(\frac{y_{i, \rm line} - m_i(\theta)}{(\kappa\sigma_i)}\right)^2\right]}{(\kappa\sigma_i)\sqrt{2\pi}}
\end{equation}
where $\kappa$ is a ``noise boost'' parameter to account for the presence of unknown systematic effects not included in the reported error bars, such as those due to flux calibration, wavelength calibration, and continuum subtraction.

Once the posterior distribution is obtained, many different algorithms are available for exploring it and computing summaries such as marginal distributions for parameters. We have implemented our model with two methods, the first is Metropolis-Hastings, a Markov-Chain Monte Carlo (MCMC) algorithm, which provides samples from the posterior PDF for the model parameters.  The second is Diffusive Nested Sampling \citep{dnest}, which provides samples from the posterior PDF {\it and} an estimate of the evidence value for the model.  Although the evidence calculation makes the second algorithm significantly slower than the first, Diffusive Nested Sampling is much faster than alternative MCMC-based implementations of Nested Sampling \citep{dnest}.  The results presented here to test the method use the MCMC algorithm, while the Diffusive Nested Sampling algorithm is used to apply the method to real reverberation mapping data \citep[][in prep]{brewer11}.

\subsection{Continuum Interpolation}
\label{sect_contint}
In order to create a mock line flux time series to compare with the data, it is necessary to interpolate between the continuum flux datapoints. Linear interpolation is the simplest approach, but it does not provide an estimate of the uncertainty in the interpolation, suggesting that we know precisely the value of the continuum $f(t)$ at all times between the measured datapoints. If we want to obtain reliable uncertainties in our results, we should acknowledge the uncertainty introduced by the interpolation process.

To account for this, we consider the entire continuum function $f_{\rm cont}(t)$ to be an unknown parameter to be inferred from the data. The prior distribution for $f_{\rm cont}(t)$ is a Gaussian Process \citep{2003itil.book.....M, rasmussen}, which is a convenient class of probability distributions over function space. Given a mean function $\mu(t)$ and a covariance function $C(t_1, t_2)$, the probability distribution for the function value $f$ at any finite set of times is a multivariate Gaussian:
\begin{equation}
p(\mathbf{f}|\mu, C) = \frac{1}{\sqrt{(2\pi)^n {\rm det }\mathbf{C}}} \exp \left(-\frac{1}{2}(\mathbf{f} - \bm{\mu})^T\mathbf{C}^{-1}(\mathbf{f} - \bm{\mu})\right)
\end{equation}
where $\mathbf{\mu}$ is a vector of means at the relevant time-points, and $\mathbf{C}$ is the covariance matrix, obtained by evaluating the covariance function at the relevant times. In the reverberation mapping problem, $f_{\rm cont}(t)$ is constrained by {\it two} data sets: the continuum measurements, and the line measurements. We parameterize the covariance function and mean function with four hyperparameters: $\mu$ (the long-term mean), $\sigma$ (the long-term standard deviation), $\tau$ (typical timescale of variations) and $\alpha$ (a smoothness parameter between 1 and 2), such that the mean function is a constant $\mu(t) = \mu$ and the covariance function is
\begin{equation}
C(t_1, t_2) = \sigma^2 \exp\left[-\left(\frac{|t_2 - t_1|}{\tau}\right)^\alpha\right]
\end{equation}

The posterior distribution function for $f(t)$ given some continuum data (but not the line data) is shown in Figure~\ref{continuum}. Note that outside the areas where we have data, the uncertainty gets large, but in areas where the data are well sampled, the uncertainty in the interpolation is small. We keep track of $f(t)$ at 500 times, both slightly preceding and following the data. Further interpolation between these 500 points is linear.  500 continuum parameters is sufficient to render the distance between continuum flux points much smaller than the maximum monitoring cadence, allowing us to resort to linear interpolation only on scales not probed by the data.  We change the 500 parameters in the same way as the model parameters, with every new proposal for the continuum function related to the one before. The function $f(t)$ can be parameterized by 500 variables with standard normal priors, which are converted to $f(t)$ values by multiplication with the Cholesky decomposition of $\mathbf{C}$. We note that our Gaussian Process method for interpolation, in the special case $\alpha=1$, is equivalent to the method of \citet{zu10}, apart from computational details. $\alpha=1$ has also been used in detailed studies of quasar variability \citep[e.g.][]{2010ApJ...721.1014M}.

\subsection{Creating Mock and Simulated Data}
Given the phase-space density for the BLR gas and the continuous continuum light curve, we can easily create a mock line flux timeseries by adding together the line flux from all the gas, which is proportional to the continuum flux at the respective lag of the gas.  The resulting mock line flux timeseries can then be compared to the reverberation mapping data and does not depend on the kinematics of the gas.  If we include the velocity information of the gas, we can create a mock spectrum for each point in the timeseries.  In order to create a mock spectrum, we make a histogram weighted on flux of the amount of gas with a given velocity using the same velocity resolution as the data.  We then convolve the histogram with a gaussian whose width is defined by a combination of thermal broadening and instrumental resolution.  The mock spectrum can then be compared to the reverberation mapping spectral data and depends on the kinematics of the gas.

\begin{figure}[h!]
\begin{center}
\includegraphics[scale=0.9]{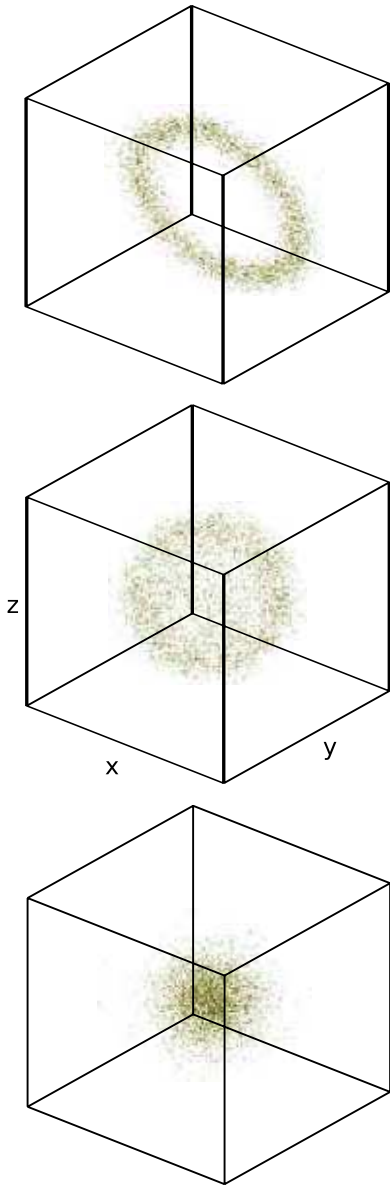}
\end{center}
\caption{Example spatial distributions of the broad line emitting gas
that can be recovered by our generic geometric model. They include a
ring/disk (top panel), a spherical shell (middle panel), and a
spherical gaussian distribution (bottom panel). \label{fig_geomodels}  }
\end{figure}

\section{Illustration and Tests Using Simple Models}
\label{sect_tests}
In order to illustrate our method, we have developed simple models of the BLR region geometry and dynamics.  As this method is fully general, it is also possible to implement more complex models within the framework described so far.  We showcase these simple models by creating simulated data with known true parameter values in our models.  This allows us to test our code as well as to explore the accuracy and precision of the results obtainable by this method of reverberation mapping analysis.  Such tests on simulated data also allow us to ascertain the data quality needed to perform inferences regarding increasingly complicated model parameters.  We showcase both geometry-only and geometry plus kinematics models, where the latter are the same as the first with the addition of velocity information given to the BLR gas.  We show transfer functions for the geometry models and velocity-resolved transfer functions for the kinematics models.  

To ensure that in our method the true parameter values are recovered, we save instances of each model and use them as simulated reverberation mapping data, adding noise and varying the timeseries characteristics to match reverberation mapping campaigns of varying quality.  A simulated dataset consists of line flux and continuum flux measurements.  Given a BLR model, the continuous continuum light curve is all that we need to create, since the mock line flux measurements can be obtained from the model and continuous continuum light curve.  We create continuous continuum light curves by using the hyperparameters of the Gaussian Processes continuum interpolation.  The hyperparameters contain information about the timescales and levels of variability in an AGN continuum timeseries.  We use values for the hyperparameters from interpolation of the Lick AGN Monitoring Project \citep[LAMP;][]{walsh09,bentz09} continuum timeseries of Arp 151, one of the most variable AGN in the LAMP sample.  The values used for the hyperparameters were $\mu = 75$ (arbitrary units), $\sigma = 30$ (same units as $\mu$), $\tau = 6 \times 10^6$ seconds and $\alpha$ = 1.5 (dimensionless).

\subsection{Geometry Model: Ring/Disk/Shell}

\subsubsection{Model definition}

We use a flexible geometry model of the BLR gas density to test our method when only integrated line flux measurements are used instead of the full spectral shape.  The model is that of a spherical shell centered on the central engine with parameters allowing partial, axisymmetric illumination of the shell and varying inclination of the resulting ring/disk.  Examples of possible configurations, ranging from a complete shell to a thin ring/disk, are shown in Figure~\ref{fig_geomodels}.  The parameters of the model are the mean radius of the disk, $r_{0}$, the thickness of the disk in the radial direction, $\sigma_r$, the illumination angle of the shell, and the inclination of the shell.  The illumination angle is defined so that values approaching 0 define an increasingly thin ring/disk and a value of $\pi/2$ defines a spherical shell.  The inclination angle is defined so that values approaching 0 define a face-on ring/disk and a value of $\pi/2$ is an edge-on ring/disk.  We use a normal distribution to define the radial thickness of the shell, so that $r_{0}$ and $\sigma_r$ are the average and 1$\sigma$ width of a normal distribution.  The normal distribution is created in the $x$, $y$, and $z$ cartesian coordinates.  

It is important to set appropriate prior probability distributions for each model parameter.  For parameters where we know the order of magnitude of the parameter value we use a flat prior in the parameter.  Examples of parameters with flat priors in the parameter include the inclination angle and the illumination angle, which may only vary between 0 and $\pi/2$.  For parameters where we do {\it not} know the order of magnitude of the parameter value we need a prior that treats many orders of magnitude equally, so we use a flat prior in the log of the parameter.  Examples of parameters with flat priors in the log of the parameter include $r_{0}$ and $\sigma_r$.  These choices of prior probability express complete ignorance in the value of a parameter within some reasonable range, but it is necessary to make the distinction between whether or not the order of magnitude of a parameter value is known.  In the cases considered in the remainder of this paper, the posterior is much narrower than the prior, and therefore the inference is dominated by the likelihood, i.e. the information contained in the data.

The underlying spherical symmetry of these models and the angular dependence of the ring/disk model allow us to use spherical coordinates.  In order to sample the gas phase-space density at a finite number of points, we use a grid in $\log(r)$, $\phi$, and $\cos(\theta)$.  Using equal steps in $\cos(\theta)$ instead of $\theta$ means that the volume of each grid point depends only on the radius, $r$.  The density is then multiplied by the volume of the grid point to find the total mass of gas in each grid point.  The emissivity of each grid point also depends on the radius $r$ because the continuum ionizing radiation flux falls off as $r^{-2}$, requiring more gas mass at larger radii to have the same line flux contribution as less gas mass at smaller radii.  In general, the illumination parameter allows us to model any axisymmetric ionizing flux.

We test our method to recover the BLR model parameters by creating simulated data, the true parameter values of which are given in Table~\ref{table_simdata}.  The continous continuum function is obtained using the hyperparameters from the Gaussian Processes interpolation of Arp 151 reverberation mapping data, as described in Section~\ref{sect_contint}, and evaluated at 120 consecutive ``observations'' one day apart.  The line flux timeseries for each model are generated using this continuous continuum function and a given set of model parameters.  The line flux timeseries contain 60 ``observations'' one day apart, starting 60 days after the start of the continuum flux ``observations''.  These simulated data are meant to represent excellent reverberation mapping data, with an observation campaign of similar length to recent campaigns \citep[see e.g.][]{bentz09}, but without gaps due to difficult weather conditions.  Additional noise has also been added to the simulated data.  Most simulated datasets have line flux errors of 1.5\%, which represents very favorable observing conditions, but we have also tested simulated data with errors of 5\% to reflect the current typical error of reverberation mapping line flux measurements.

\begin{figure}[h!]
\begin{center}
\includegraphics[scale=0.3]{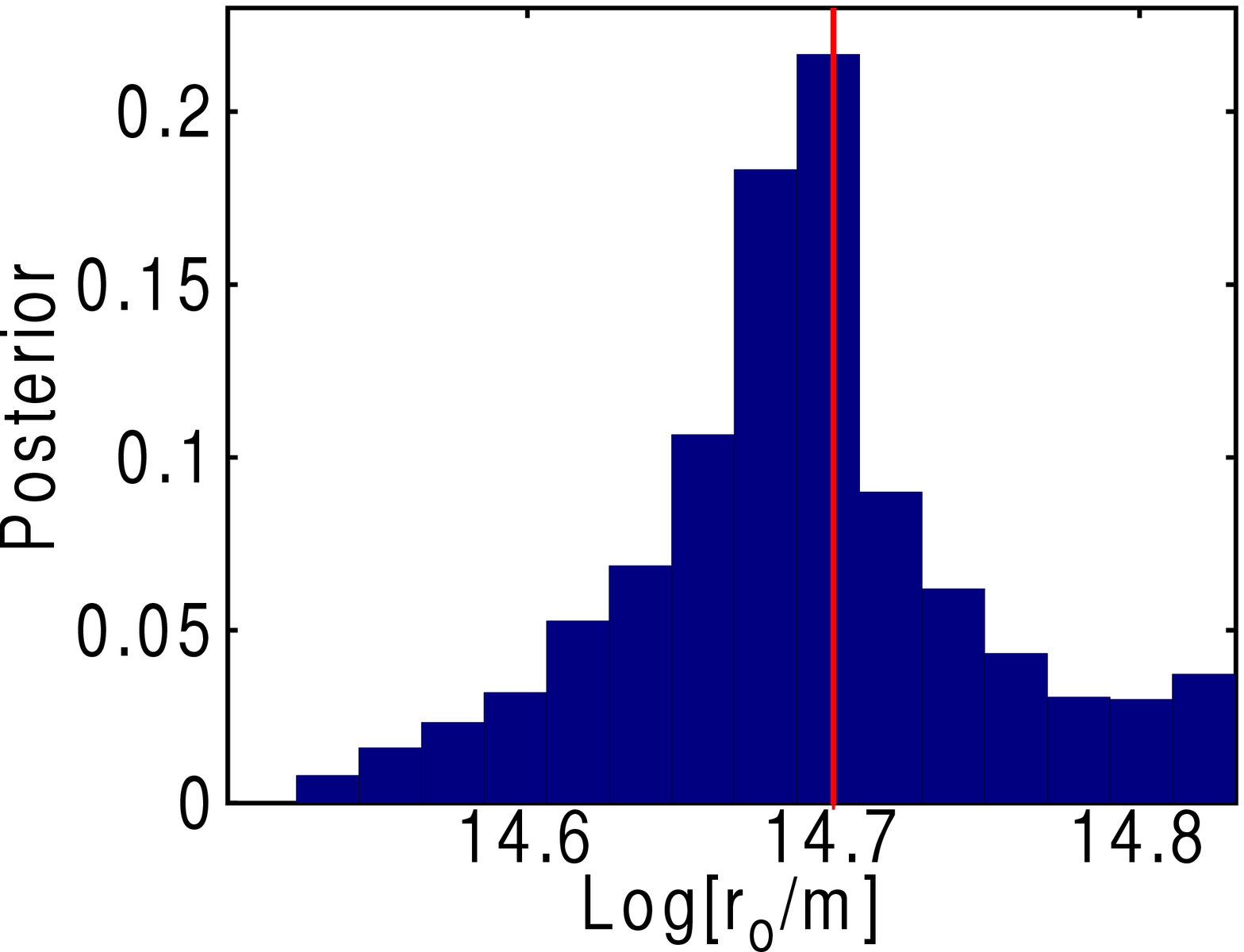}
\includegraphics[scale=0.3]{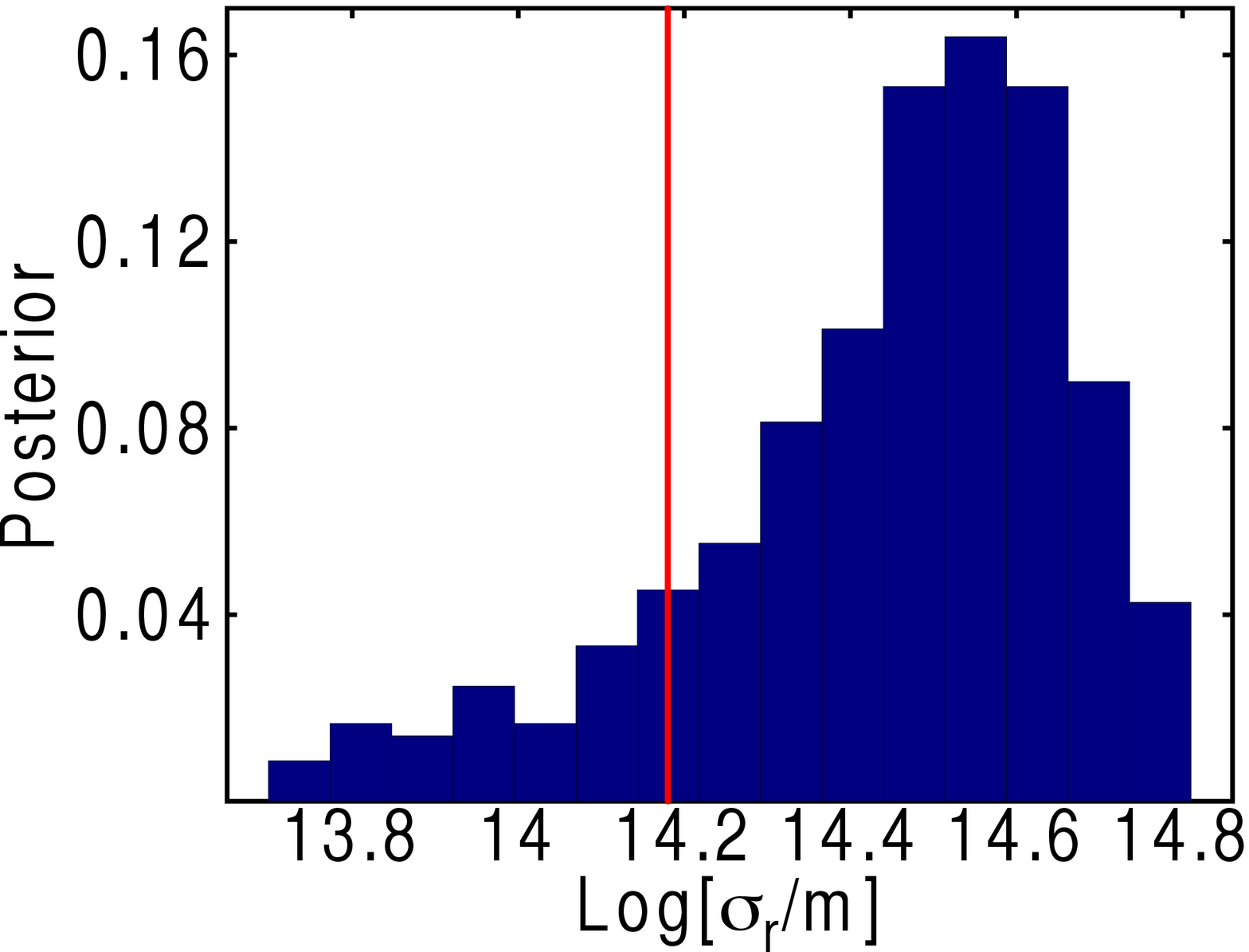}
\includegraphics[scale=0.3]{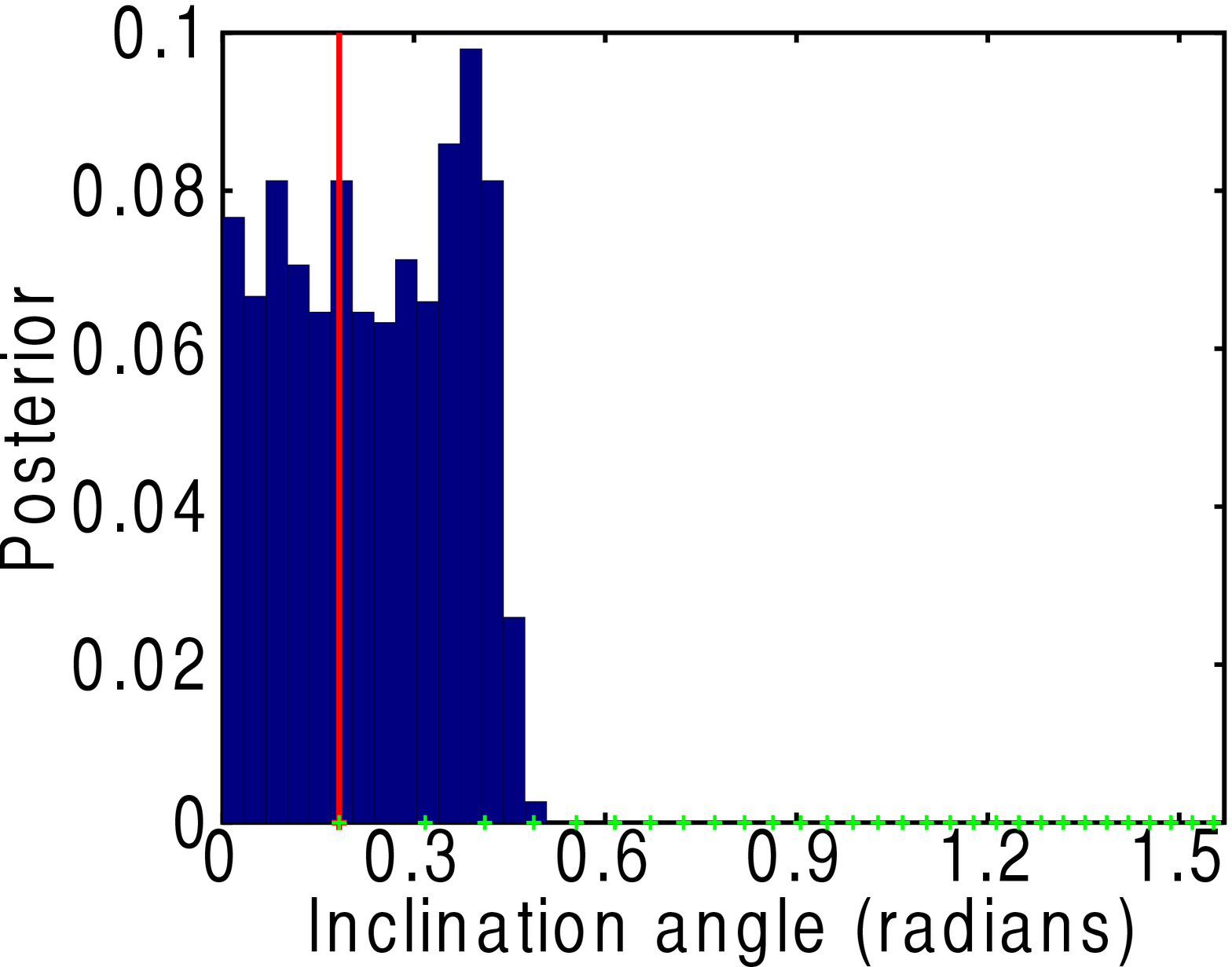}
\includegraphics[scale=0.3]{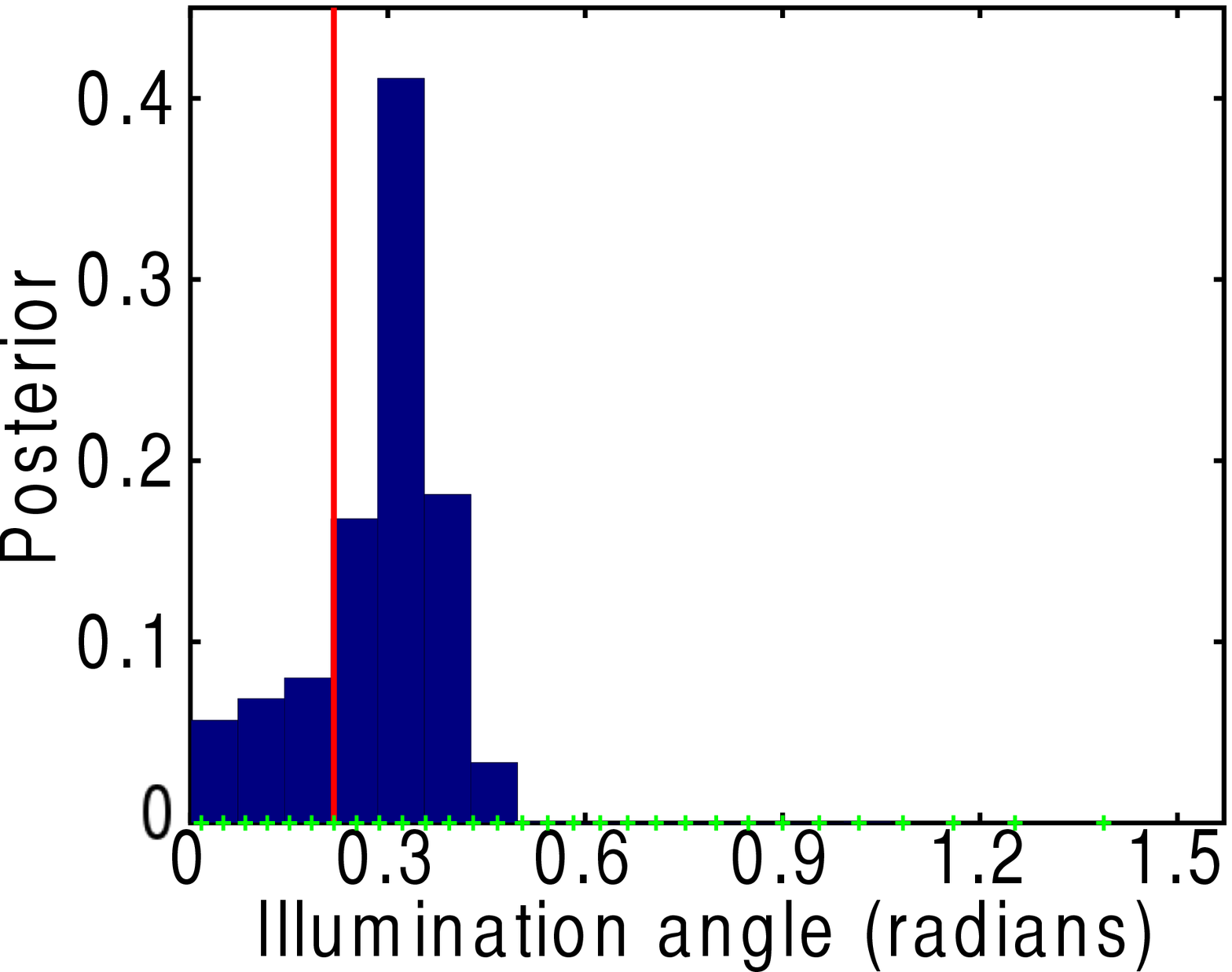}
\end{center}
\caption{Posterior probability distributions for face-on disk geometry model parameters of simulated data 4 (see Table \ref{table_simdata}) with 1.5\% line flux uncertainty.  Top to bottom: $r_{0}$, $\sigma_r$, inclination angle, and illumination angle.   The inclination angle and illumination angle both have a resolution given by the grid in $\cos\theta$.  The true value for each parameter in this model is shown by the vertical red line and the grid in $\cos\theta$ is shown along the x-axis with green crosses for the angular parameters.  The grid used to create these posterior distributions is 60 steps in $\log(r)$, 40 steps in $\phi$, and 60 steps in $\cos\theta$.    \label{fig_pdf10}}
\end{figure}

\begin{figure}[h!]
\begin{center}
\includegraphics[scale=0.3]{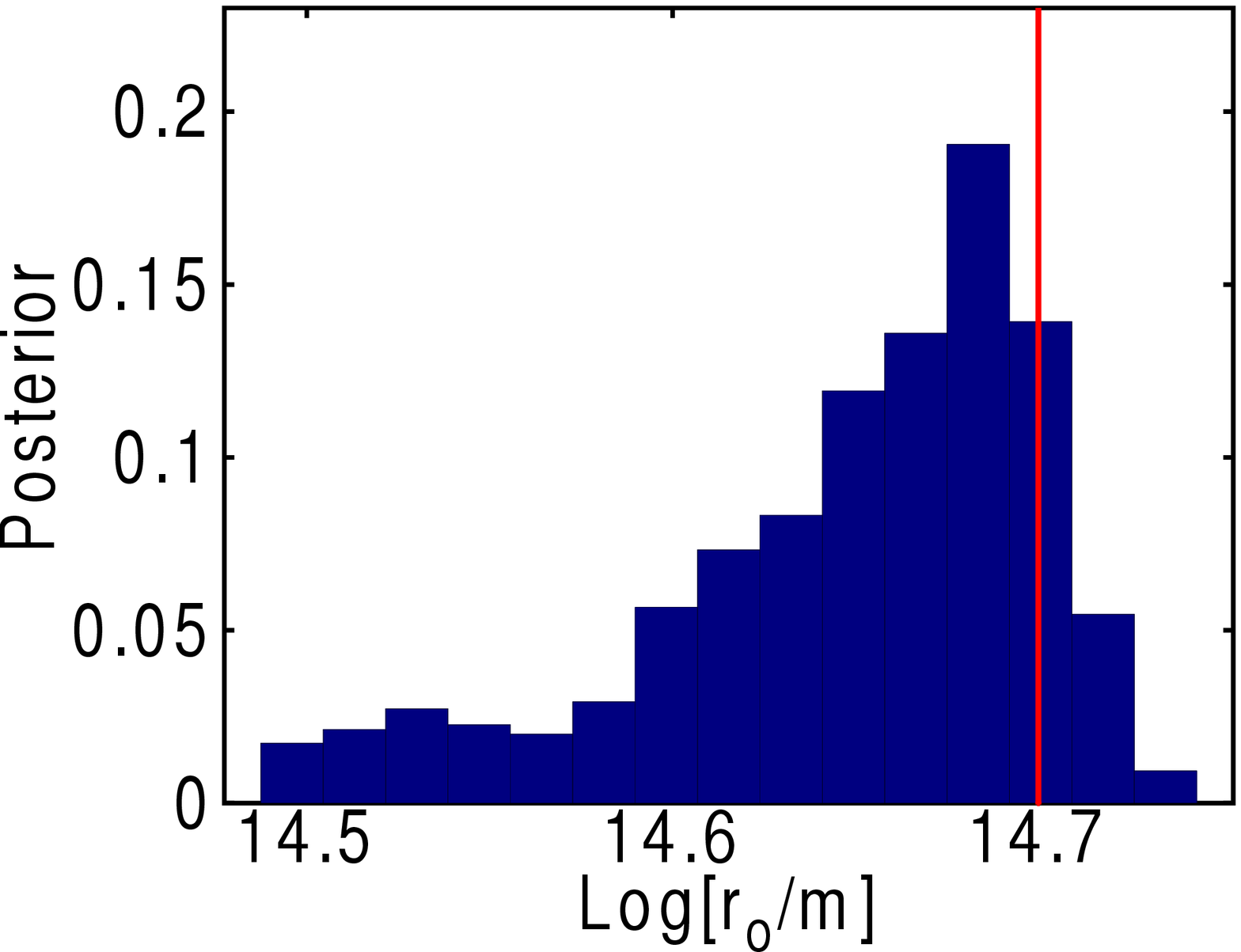}
\includegraphics[scale=0.3]{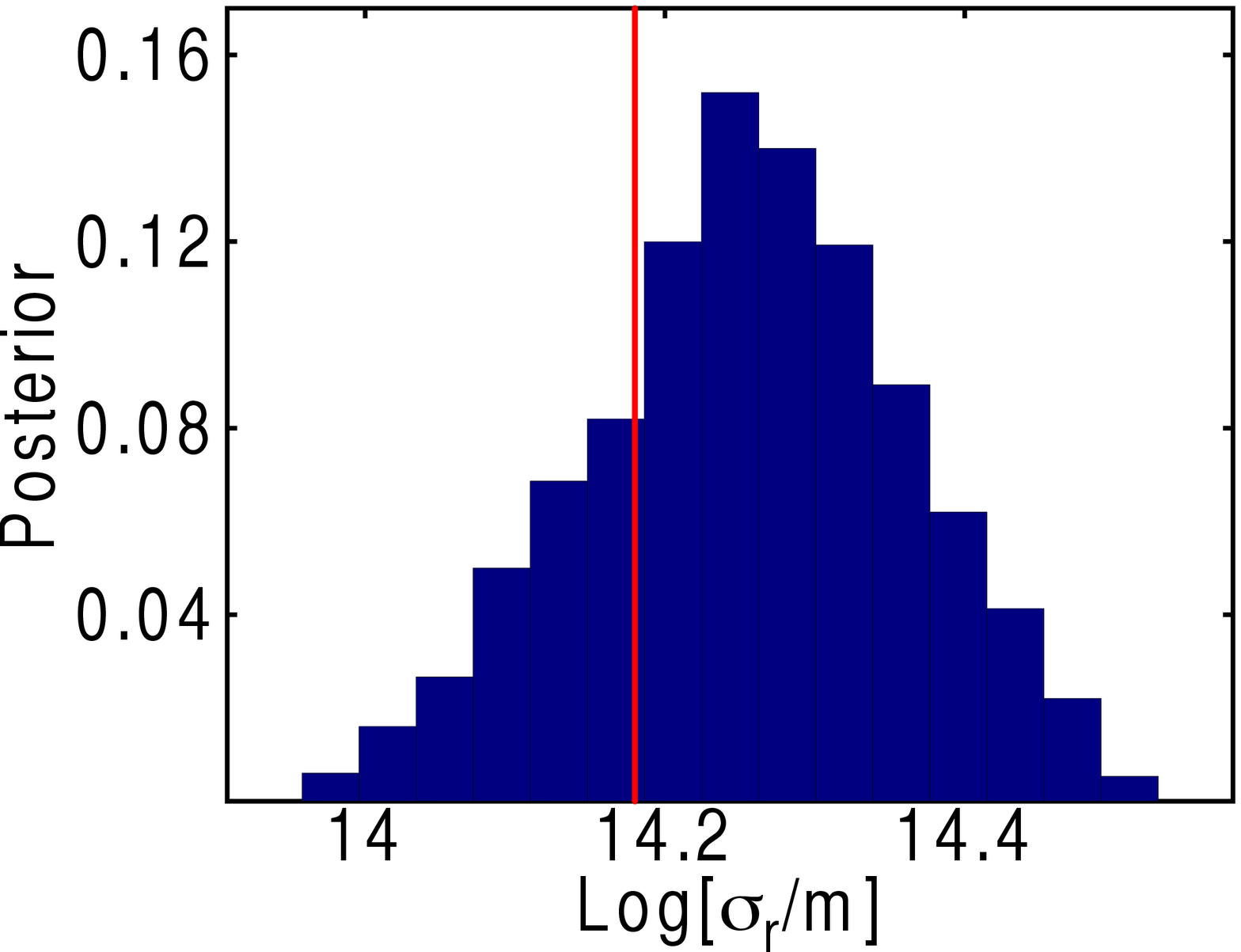}
\includegraphics[scale=0.3]{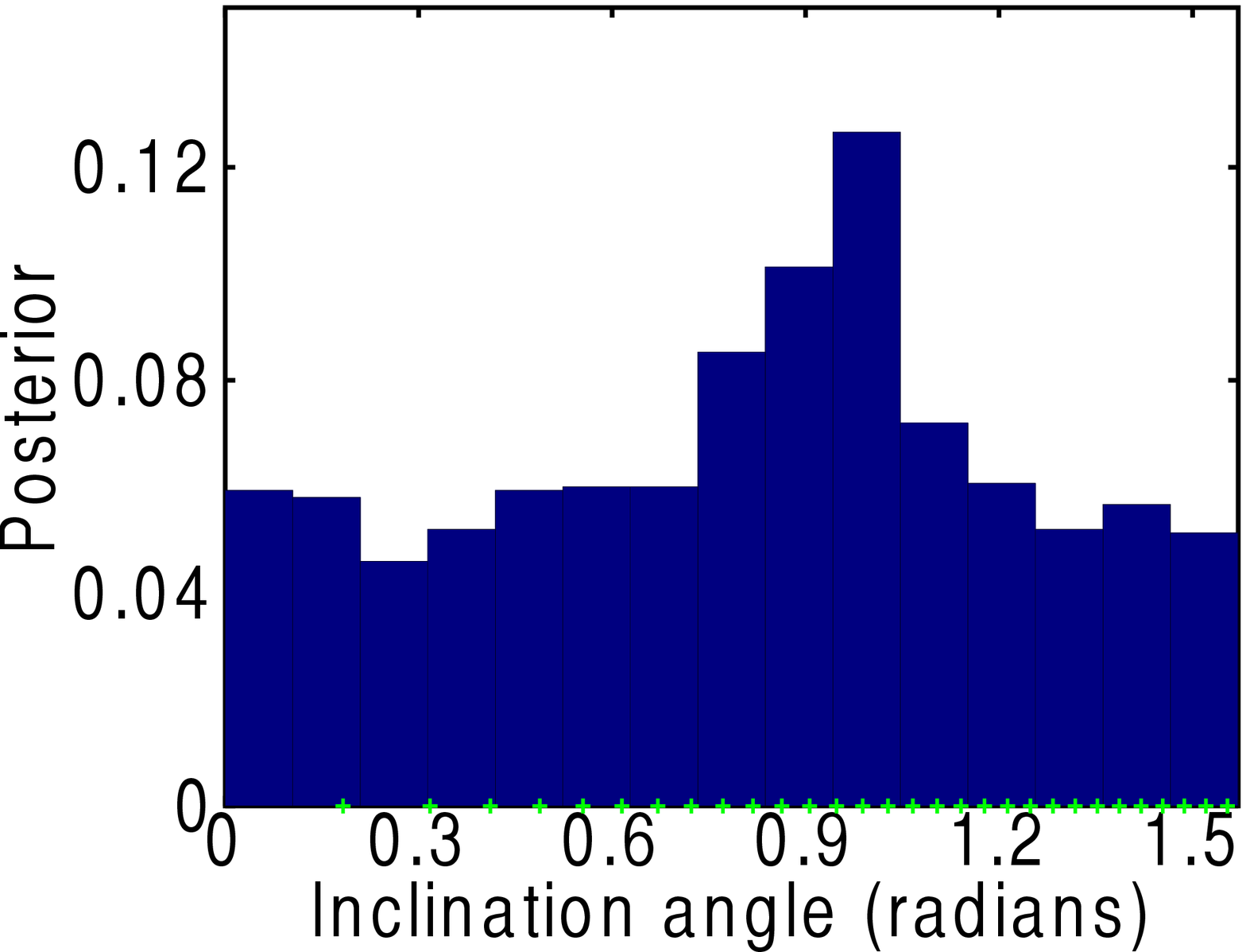}
\includegraphics[scale=0.3]{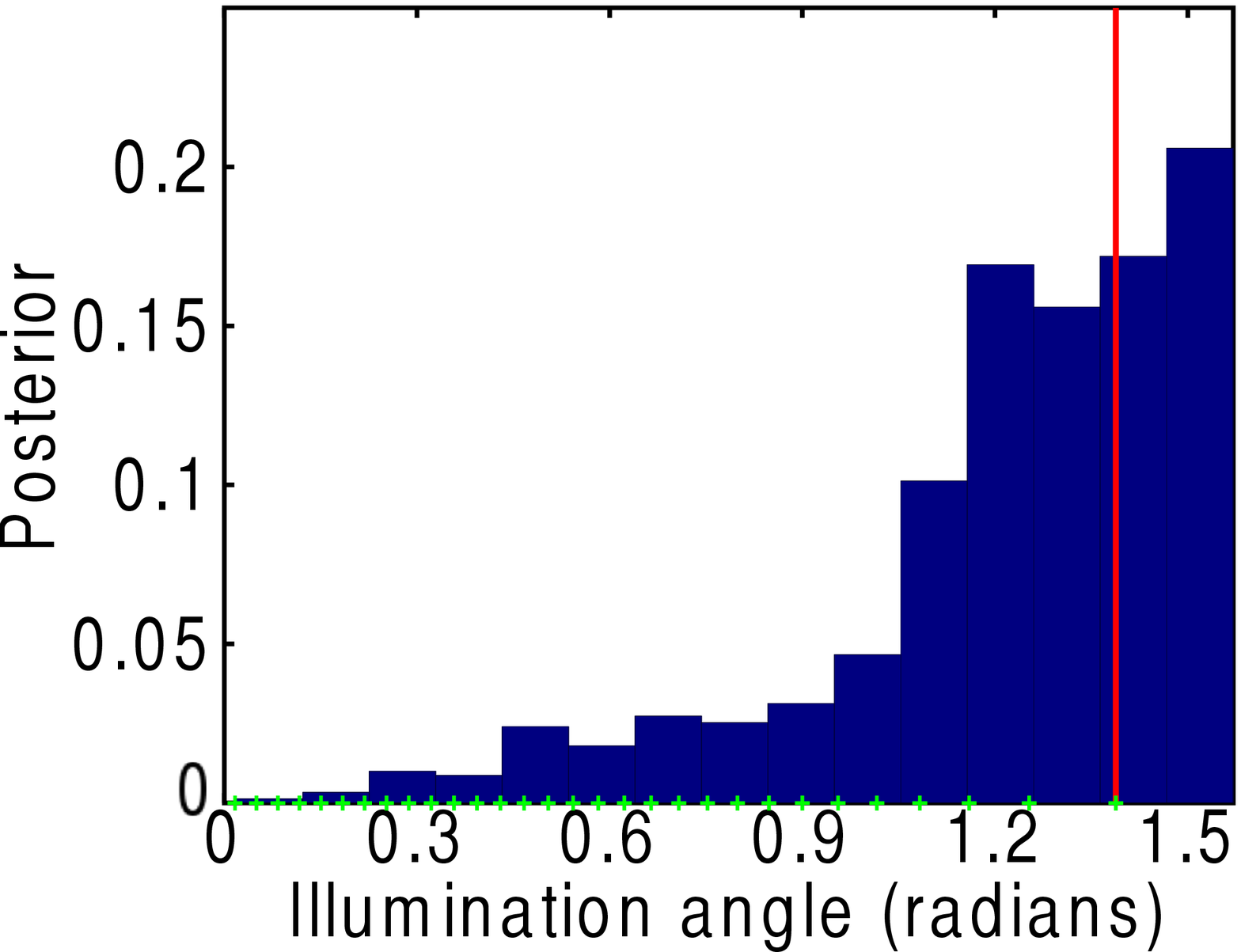}
\end{center}
\caption{Posterior probability distributions for shell geometry model parameters of simulated data 5 (see Table \ref{table_simdata}) with 1.5\% line flux uncertainty.  Top to bottom: $r_{0}$, $\sigma_r$, inclination angle, and illumination angle.  The true values for the parameters and the grid points are shown as in Figure\,\ref{fig_pdf10}.  Note that since the simulated data is spherically symmetric, it should not strongly prefer an inclination angle, and thus no true parameter value is shown in the inclination angle pdf.    \label{fig_pdf11}}
\end{figure}

\begin{figure}[h!]
\begin{center}
\includegraphics[scale=0.45]{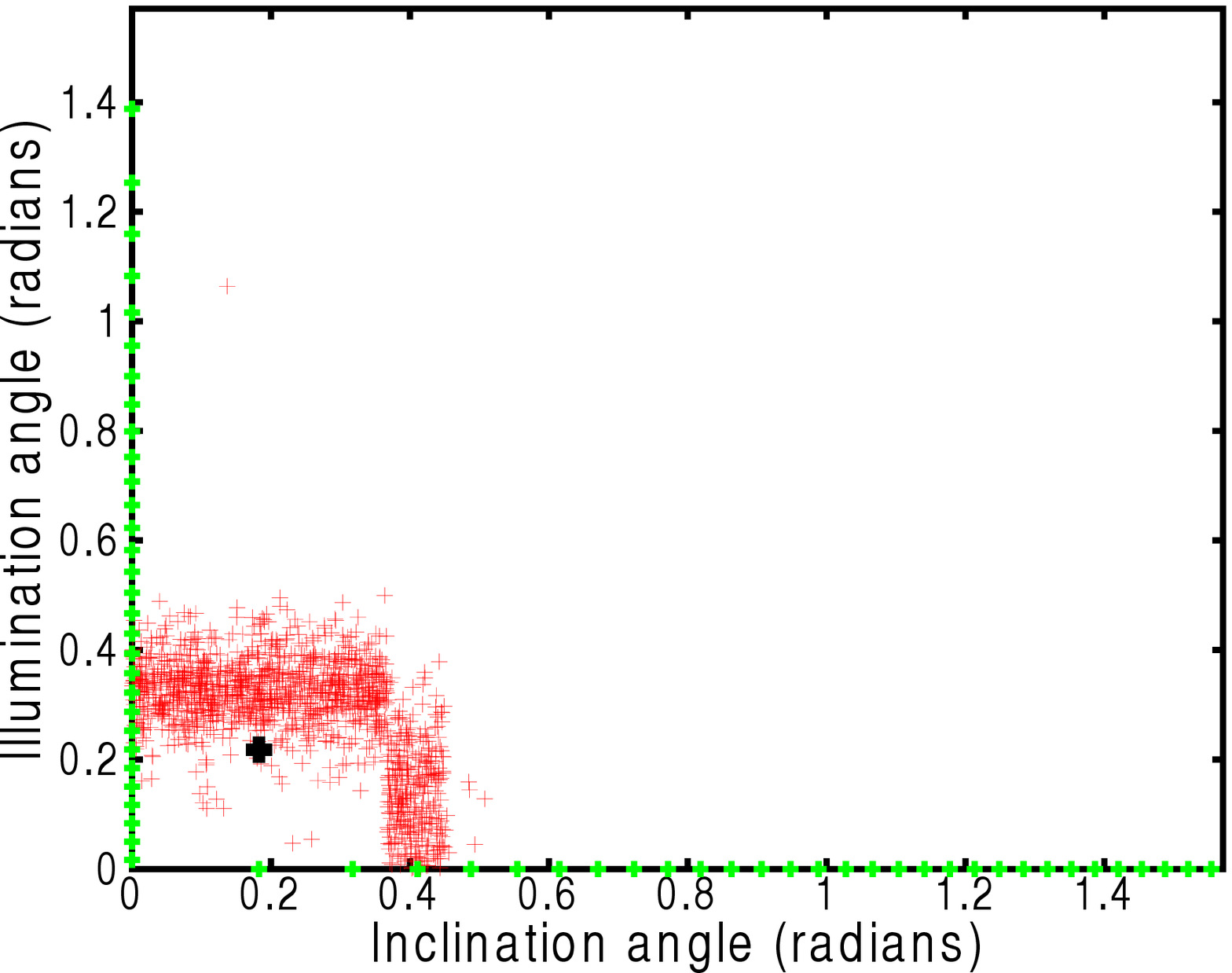}
\includegraphics[scale=0.45]{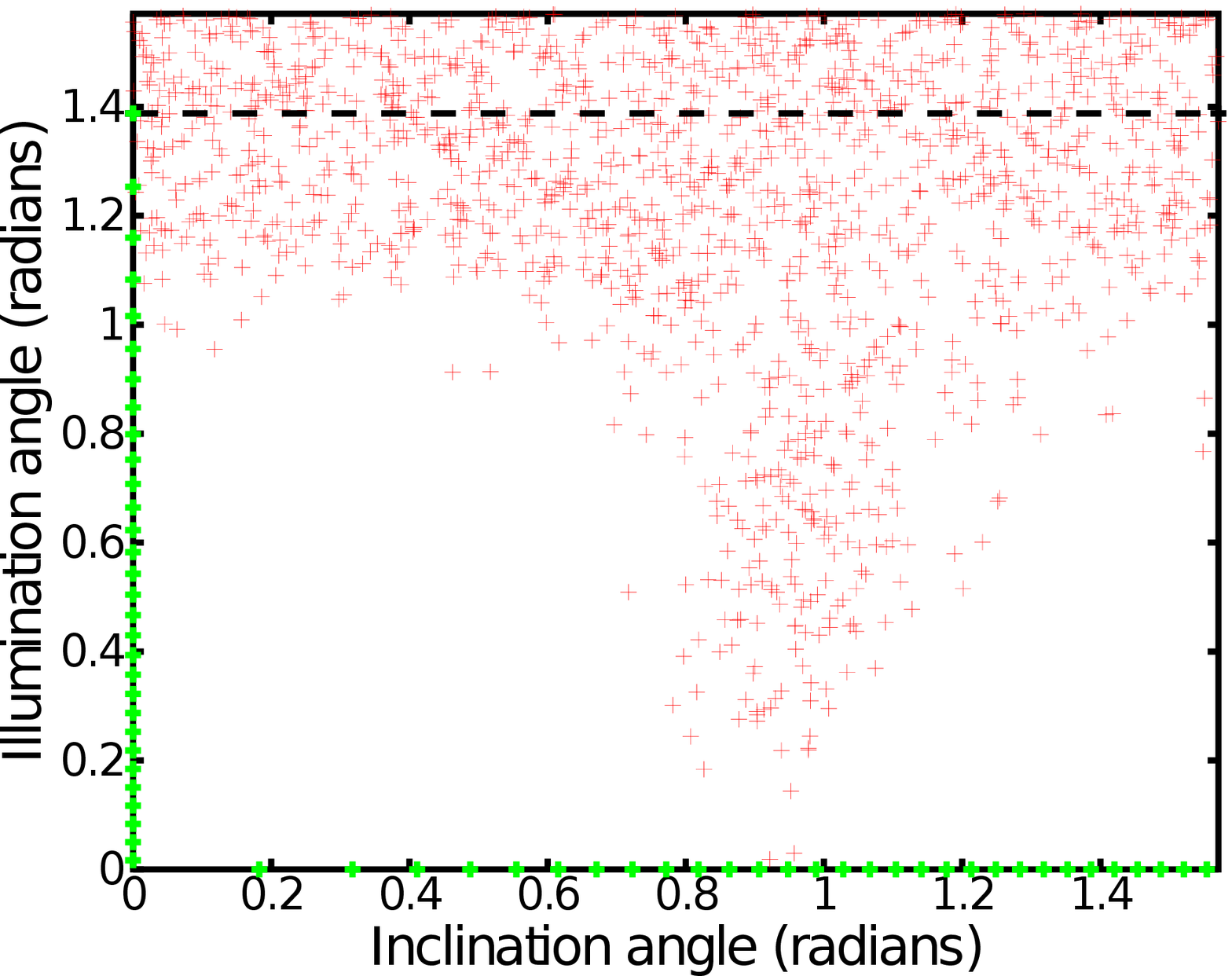}
\end{center}
\caption{Joint posterior probability distributions for inclination and illumination angles for face-on disk with 1.5\% line flux uncertainty (simulated data 4) and shell with 1.5\% line flux uncertainty (simulated data 5).  The true parameter values are shown by (top) the black cross and (bottom) the black dashed line.  \label{fig_pdf}}
\end{figure}

\begin{figure}[h!]
\begin{center}
\includegraphics[scale=0.45]{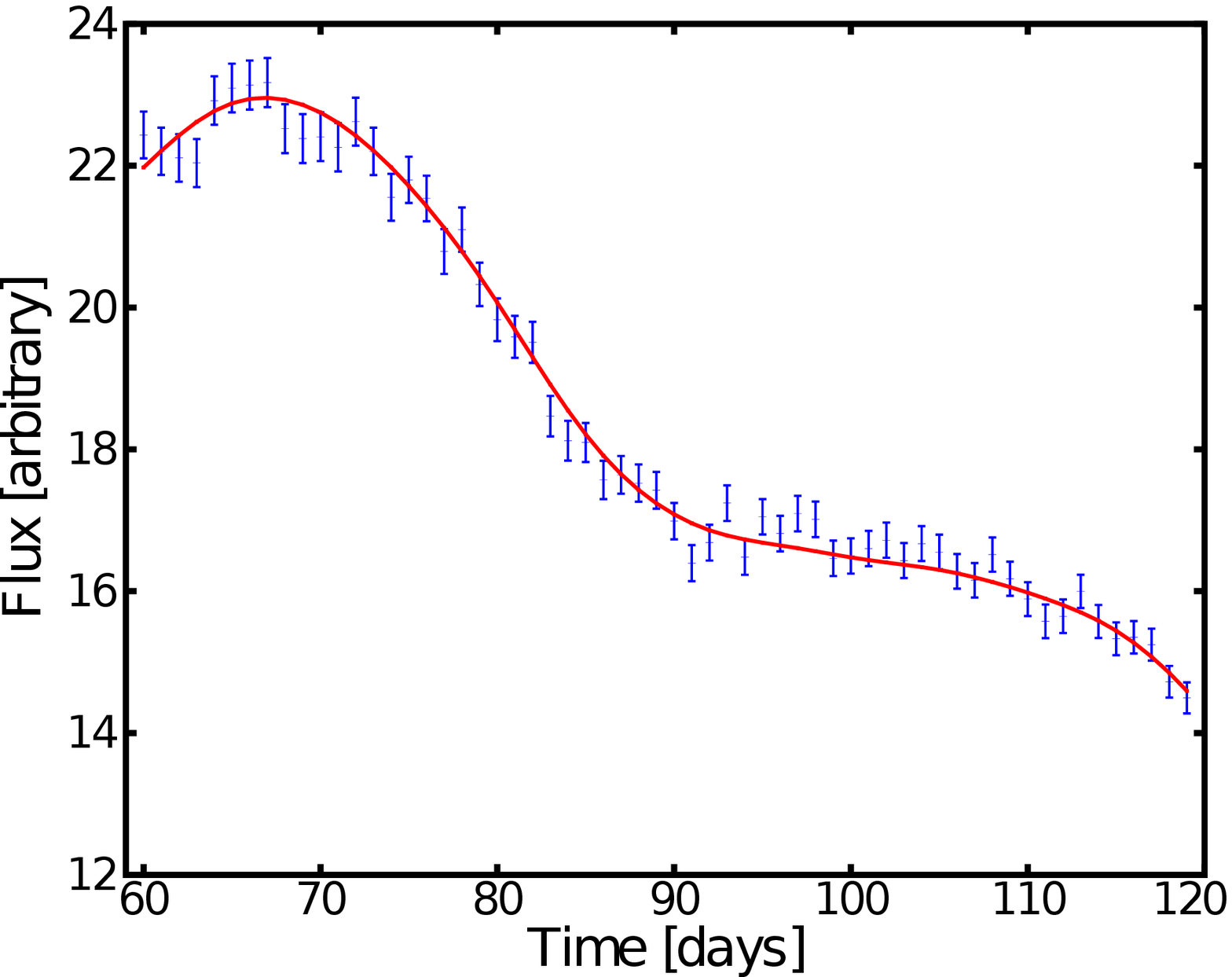}
\includegraphics[scale=0.45]{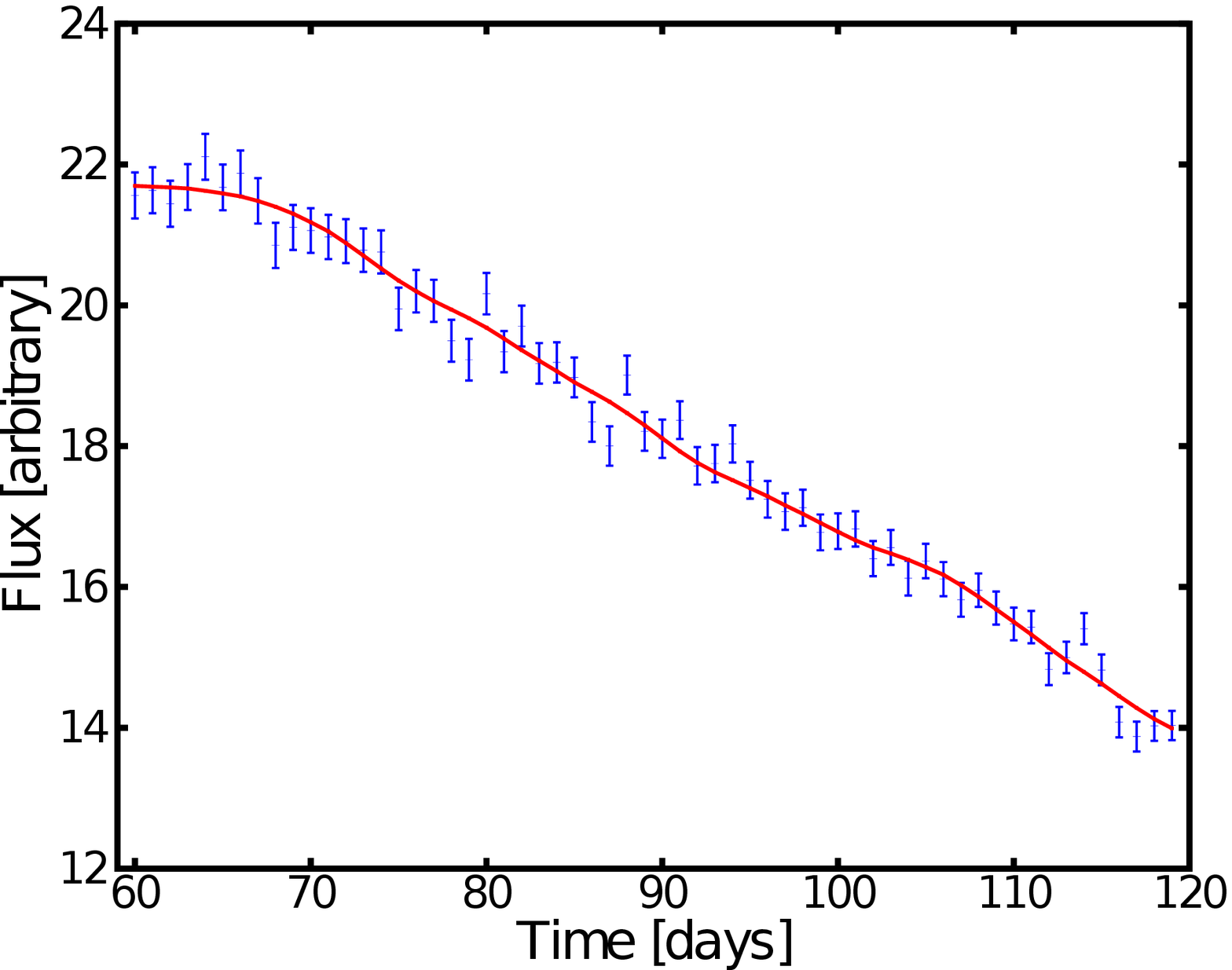}
\end{center}
\caption{Timeseries for face-on disk (simulated data 4, top panel) and shell (simulated data 5, bottom panel), both with 1.5\% line flux uncertainty.  Simulated data are shown in blue with error bars and the mock data from a random set of parameter values sampled from the posterior is shown in red.  The continuum light curve used to create these line light curves is shown in Figure\,\ref{continuum}.  \label{fig_timeseries}}
\end{figure}

\begin{figure}[h!]
\begin{center}
\includegraphics[scale=0.45]{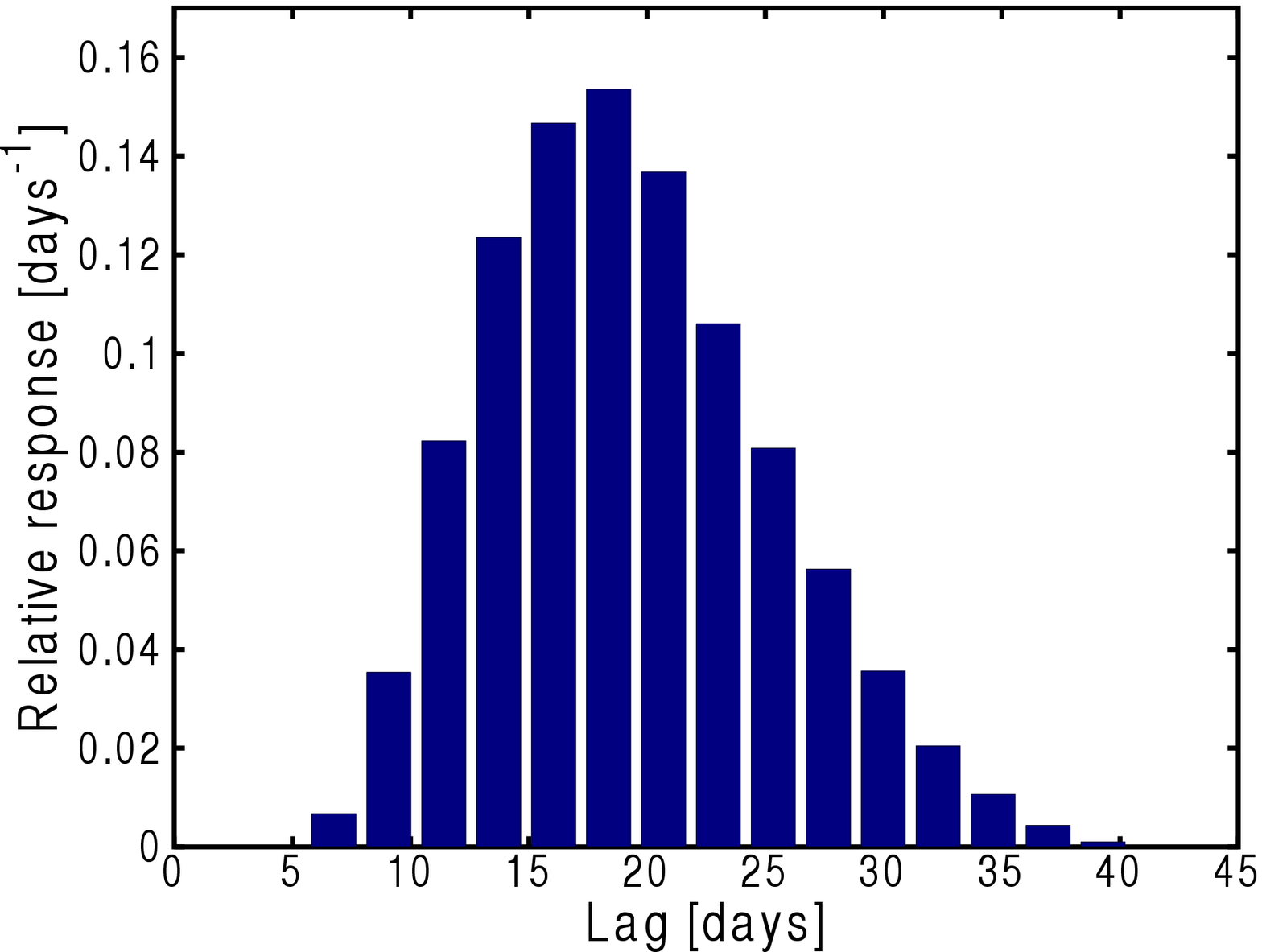}
\includegraphics[scale=0.45]{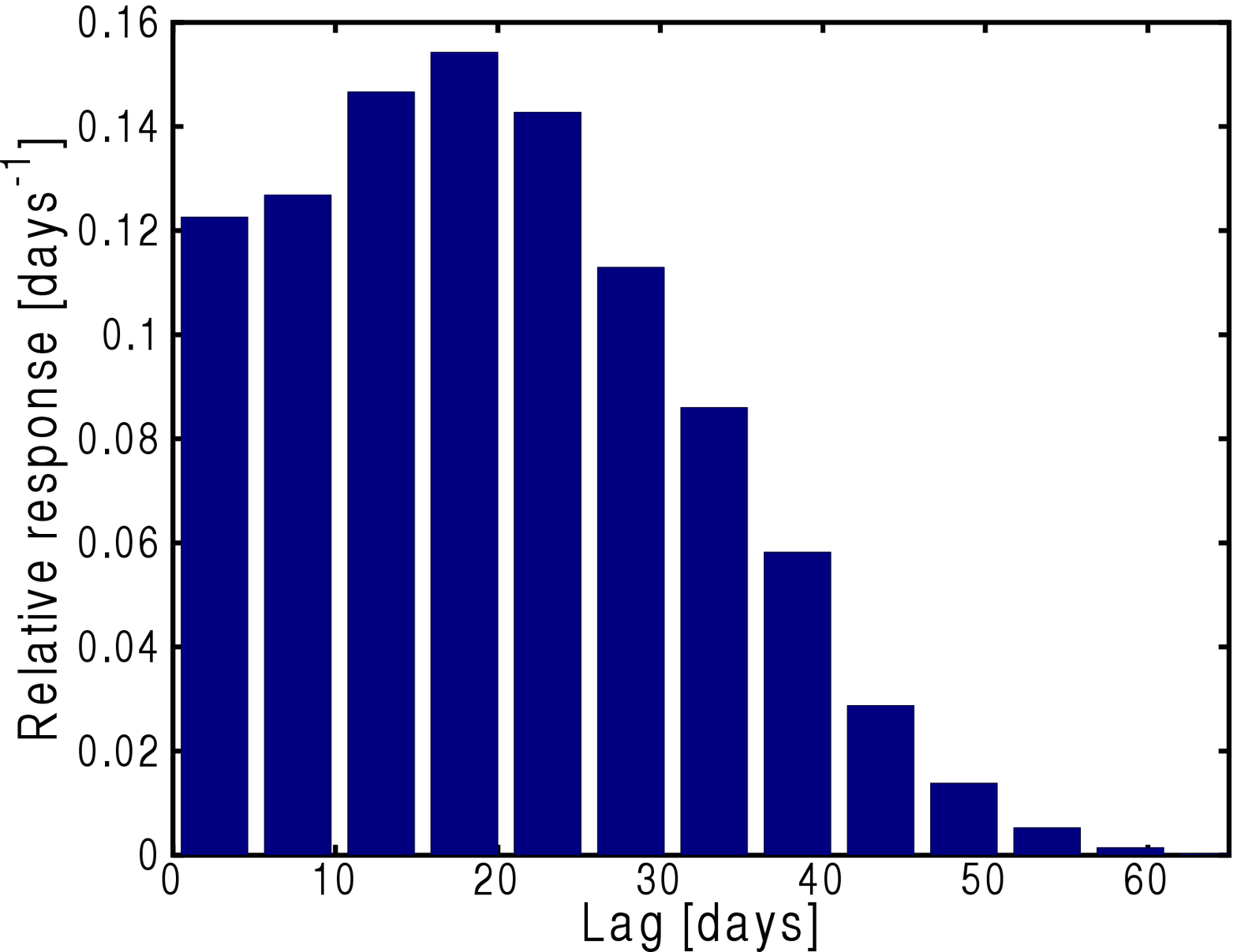}
\end{center}
\caption{Velocity-unresolved transfer functions for face-on disk (simulated data 4) and shell (simulated data 5), both with 1.5\% line flux uncertainty.  The same grid was used to make these transfer functions as was used to obtain the posterior probability distributions shown in Figures\,\ref{fig_pdf10} and \ref{fig_pdf11}.  \label{fig_transfer}}
\end{figure}

\subsubsection{Testing the geometry model}

The first test is whether we can recover the parameter values of the
simulated data using the MCMC algorithm described in
Section~\ref{sect_method}.  Since our one flexible geometry model
encompasses a number of different geometries, such as a shell, thin or
thick ring or disk, we do not have to consider model selection at this
point.  We test the many possible geometries of this model by creating
five simulated datasets, whose true parameter values are given in
Table~\ref{table_simdata}.  The simulated datasets include an inclined
disk with line flux errors of 1.5\% and 5\% and an edge-on disk, a
face-on disk, and a shell with line flux errors of 1.5\%.  The MCMC
algorithm is typically run for 150,000 iterations and all parameter
values are recovered to within two standard deviations of the
posterior probability distributions of the parameters, with 10/13
recovered to within one standard deviation.  This is as expected,
since we should find the true parameter value to lie within $1\sigma$
about $\sim 68\%$ of the time and to lie within $2\sigma$ about $\sim
95\%$ of the time.  The mean and standard deviation of the posterior
distributions are given in Table~\ref{table_simresults}, with the
exception of many of the angular parameters, where the quoted
$1\sigma$ uncertainty does not adequately describe the posterior
distribution.  Part of the reason for the standard deviation of the
angular parameters not describing the posterior is due to the uneven
step size in $\theta$, so that values of the illumination angle close
to $\pi/2$ and values of the inclination angle close to 0 radians have
much poorer angular resolution.  This might lead to an angular
parameter being quoted as having a mean of 1.22 radians and a
$1\sigma$ uncertainty of 0.29, as for the illumination angle of the
Shell model simulated data, but while this uncertainty may seem large,
it corresponds to an uncertainty of only 1-2 grid points in $\theta$.
The posterior distributions for the face-on disk and shell simulated
data are shown in Figures~\ref{fig_pdf10} and \ref{fig_pdf11}.  Select
joint probability distributions between the inclination and
illumination angles are also shown in Figure~\ref{fig_pdf} in order to
show the degeneracies between different models.  In particular, for
the shell model, the inclination is not constrained unless the
illumination angle is small, or rather, unless the sphere of BLR gas
is not entirely illuminated.

The posterior pdfs for the five simulated datasets show that the
edge-on disk, face-on disk, and shell geometries allow for excellent
recovery of the parameter values with estimates of the uncertainty.
For the two inclined disk simulated datasets, there is some degeneracy
in the angular parameters, leading to large uncertainties in their
average values.  The MCMC algorithm finds a more likely geometry
configuration than the true configuration for the inclined disk
datasets, although the true configuration is still a valid possibility
with posterior local maxima at the true parameter values.  With the
increased simulated line flux error from 1.5\% to 5\% however, it
becomes increasingly difficult to recover the angular parameters, and
only the mean radius is recovered with a small enough uncertainty as
to be useful in describing the BLR.  This emphasizes the importance of
obtaining high quality line flux data in reverberation mapping
campaigns.

The timeseries and transfer functions for the face-on disk and shell
MCMC geometry model tests are shown in Figures~\ref{fig_timeseries}
and \ref{fig_transfer}, respectively.  The timeseries show the
simulated data overlaid with mock data created with parameters sampled
randomly from the posterior probability distributions.  The fit of the
mock data to the simulated data is excellent for all five models.  The
variety in the shape of the simulated data timeseries, all well-fit by
their respective models, shows that the MCMC algorithm for model
parameter value recovery is robust for a wide range of models.  The
transfer functions also show a variety of shapes.  For a thin shell
geometry, thinner than the shell of simulated dataset 5, our resulting
transfer function agrees with the analytic form of a tophat function
\citep[see][]{peterson93}.

\begin{figure}[h!]
\begin{center}
\includegraphics[scale=0.8]{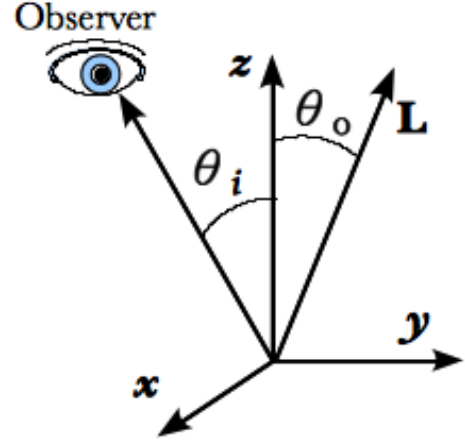}
\caption{Sketch of the dynamical model. The angular momentum vector
${\mathbf L}$ defines the plane of the orbits. Owing to cylindrical
symmetry, for each value of $\theta_0$ we consider the entire family
of ${\mathbf L}$ generated by rotation around the z-axis. The observer
is assumed to be in the x-z plane, at angle $\theta_i$ from the z-axis.
\label{kinematics_diagram}}
\end{center}
\end{figure}

\begin{figure}[h!]
\begin{center}
\includegraphics[scale=0.55]{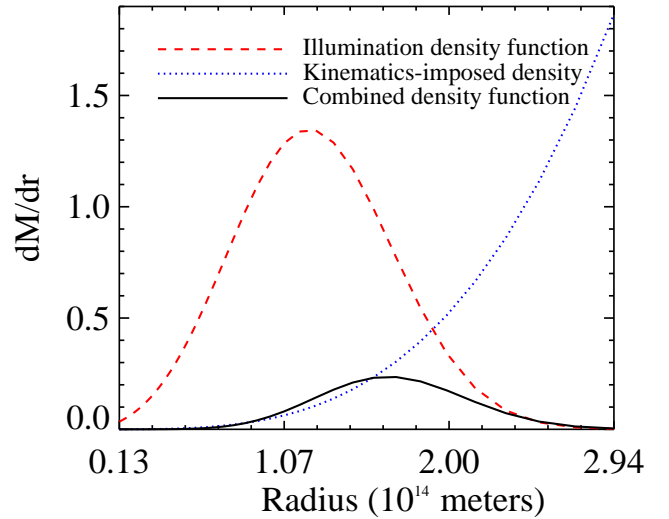}
\caption{Illustration of the combined constraints given by the
illumination function and by the dynamical model. The red line shows an
example of the distribution of illuminated BLR gas mass assuming a uniform
underlying density. The blue line shows the actual underlying mass
distribution as constrained by the dynamical model. The resulting
effective distribution of illuminated mass, consistent with both the
geometry and dynamical constraints is given by the product of the two
functions, shown in black.
\label{dynamics_illumination}}
\end{center}
\end{figure}
 
 \begin{figure}[h!]
\begin{center}
\includegraphics[scale=1.3]{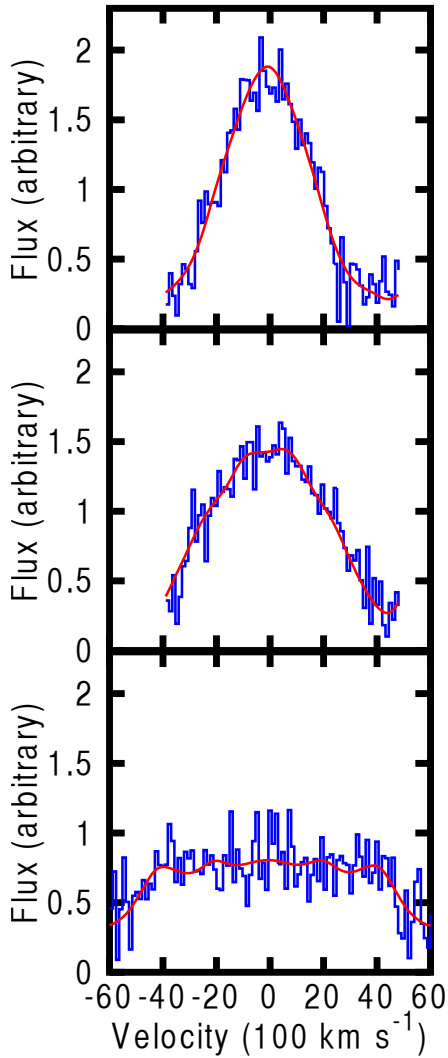}
\caption{Example spectra from three simulated datasets: (top) face-on disk with orbits confined to the disk, (middle) face-on disk with isotropic distribution of orbit orientations, and (bottom) spherical distribution with isotropic distribution of orbit orientations.  The instrumental resolution of the simulated spectra is FWHM $\sim$ 800\,km\,s$^{-1}$.  The bottom spectrum for a spherical distribution of orbits is wider than for a face-on disk because the spherical distribution allows for orbits to move directly along the line of sight, while the face-on disk only results in a small component of the BLR gas velocity lying parallel to the line of sight.  The width of the spectral line is thus directly connected to both the opening angle of the disk and the inclination angle.
\label{fig_spectra}}
\end{center}
\end{figure}

\begin{figure}[h!]
\begin{center}
\includegraphics[scale=0.4]{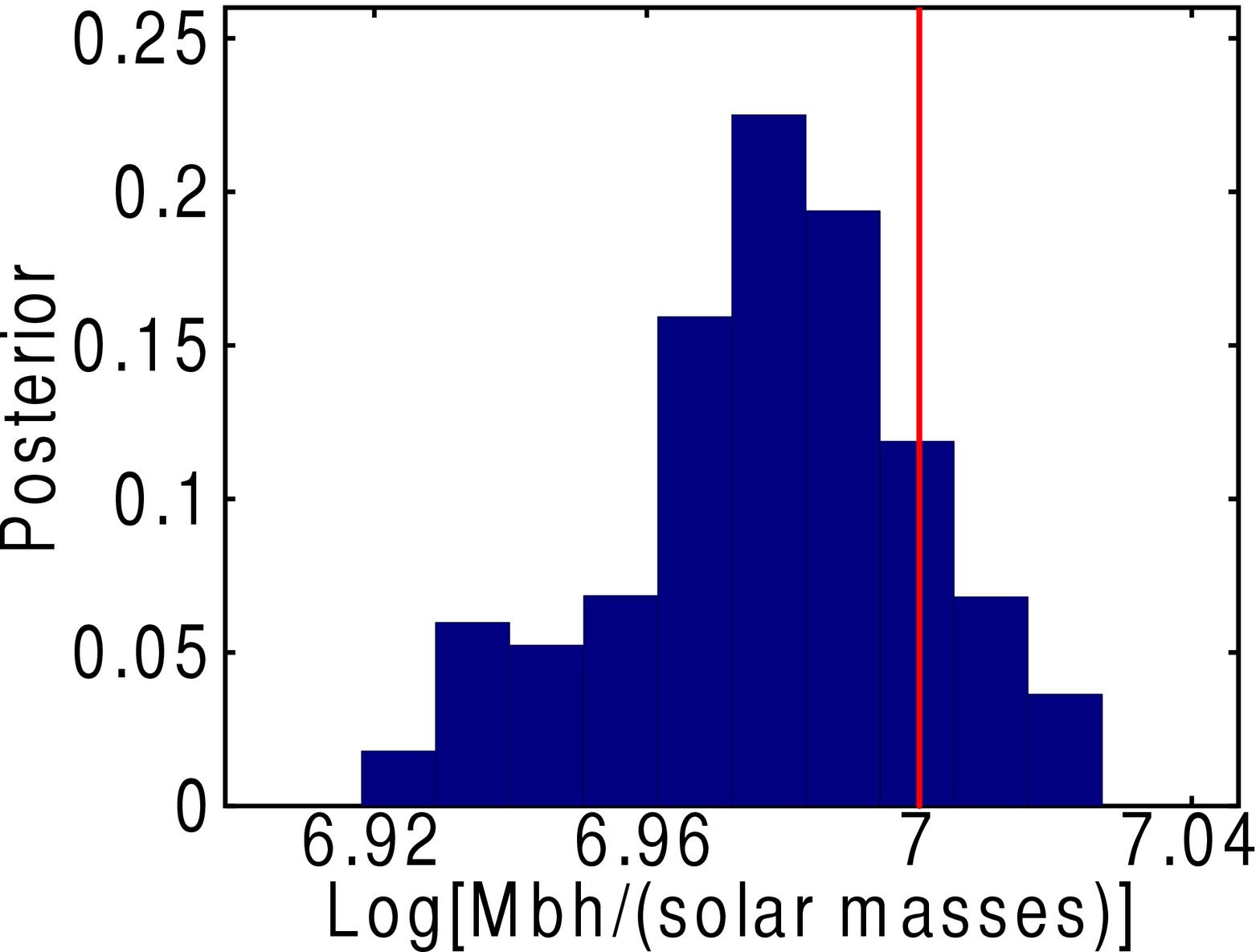}
\includegraphics[scale=0.4]{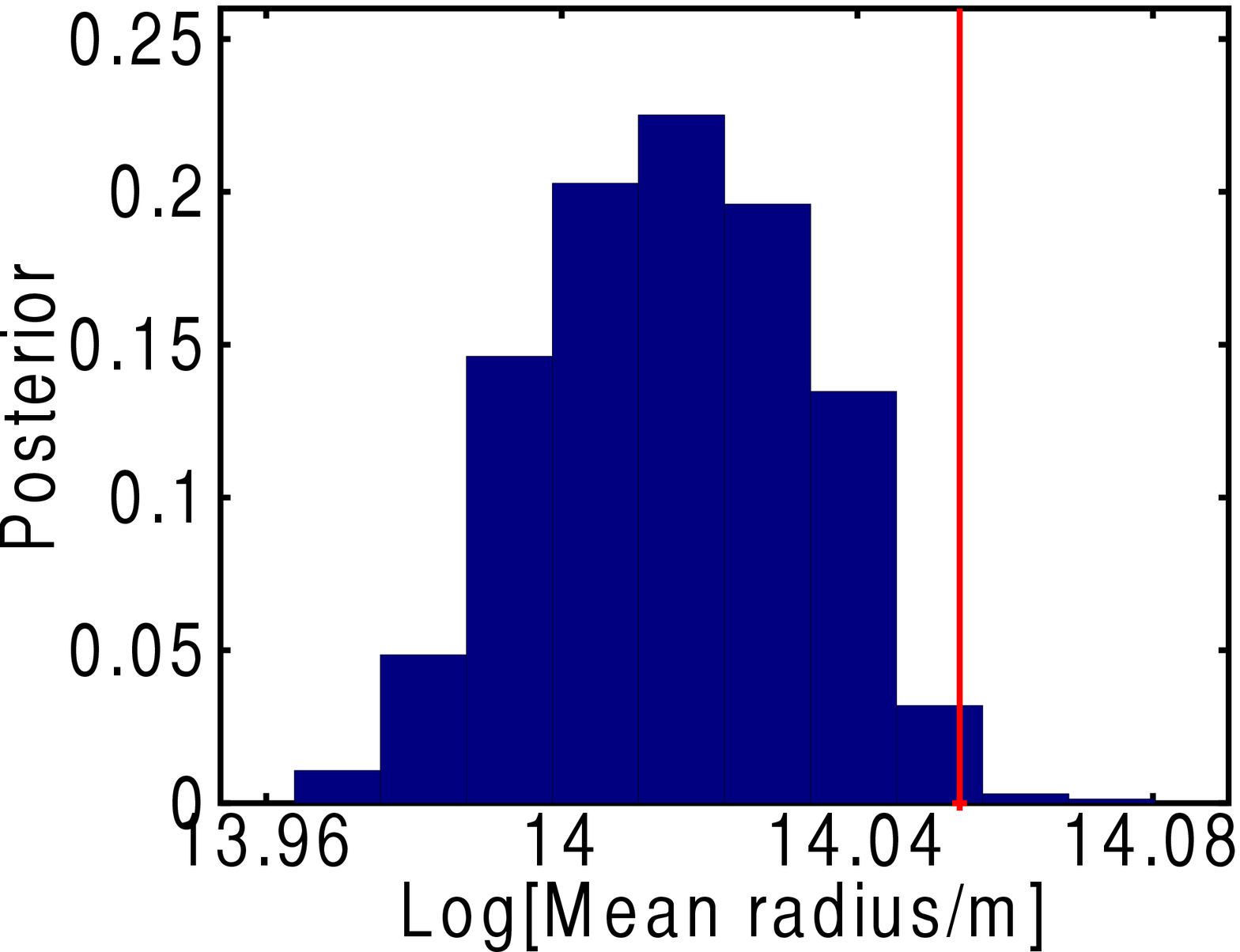}
\includegraphics[scale=0.4]{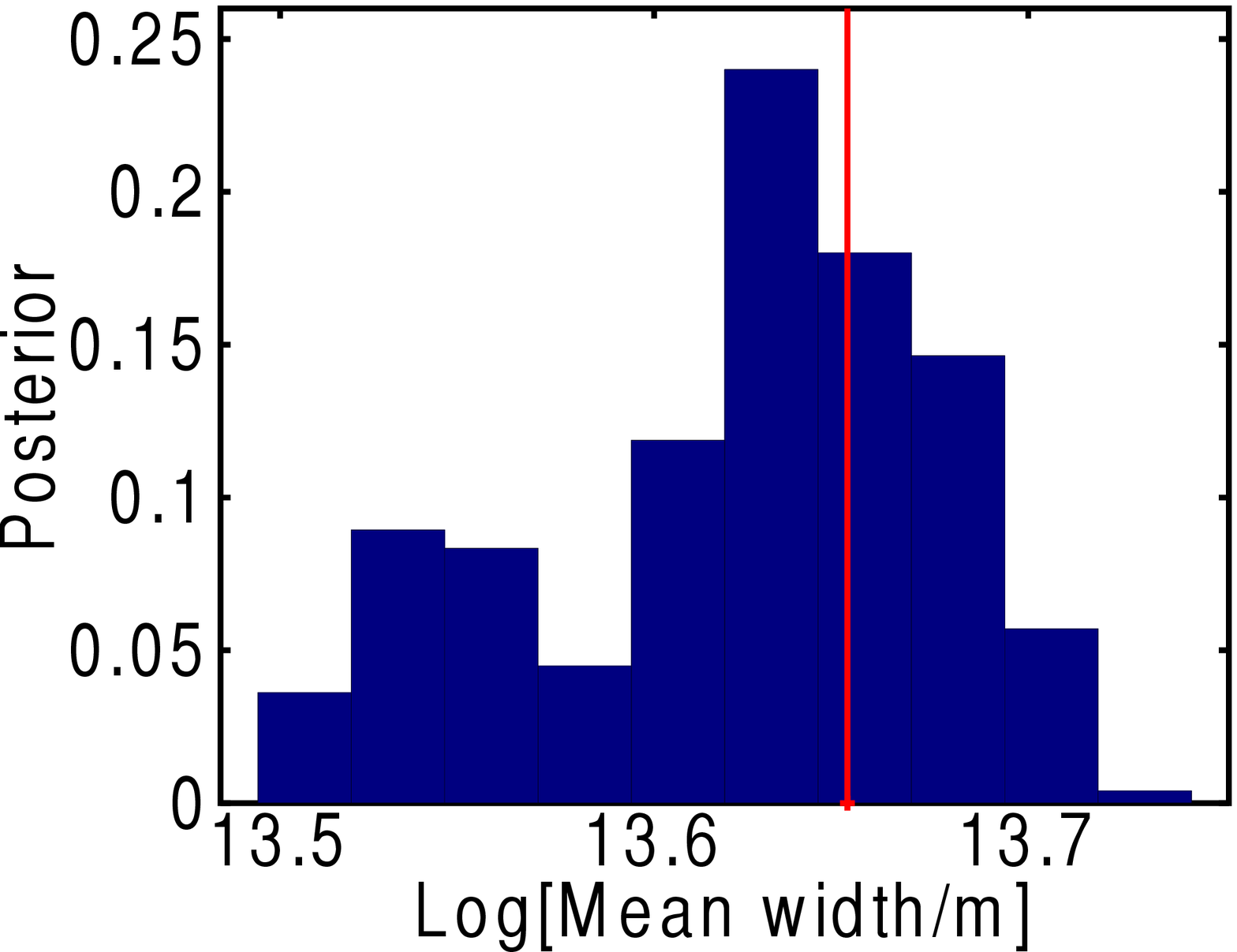}
\caption{Posterior pdfs for the first dynamical simulated dataset: face-on disk with the orbits confined to the disk.  (Top) black hole mass, (middle) the average radius of the BLR gas mass, and (bottom) the average width of the BLR gas mass.
\label{fig_dyn1a}}
\end{center}
\end{figure}

\subsection{Dynamical Model}

\subsubsection{Model definition}

In order to constrain the kinematics of the BLR and the mass of the
central black hole, we must model the velocity distribution of the BLR
gas in the context of a dynamical model. For simplicity of
illustration and speed of computation, we consider here a
cylindrically symmetric model where the BLR gas is considered to be
made of test particles in bound orbits within the spherical Keplerian
potential of the black hole.  We parameterize the
model in terms of energy and angular momentum, constants of the BLR
gas motion, so we are guaranteed velocity and geometry distributions that
do not evolve in time, and are therefore stationary during the
monitoring campaign.  In future papers we will generalize the model to
include unbound orbits to describe inflows and outflows, and also
other physical mechanisms, such as radiation pressure or
winds \citep{marconi08, netzer10}.

The model is illustrated in Figure~\ref{kinematics_diagram}. For any
choice of angular momentum ${\mathbf L}$, energy $E$ and black hole mass
$M_{\rm BH}$, the motion of the BLR test particles is then described
by the standard conservation equation resulting in elliptical orbits
in the plane perpendicular to the angular momentum. Given our
cylindrical symmetry we will consider families of angular momenta
obtained by rotation around the z-axis and defined by the polar angle
$\theta_0$ (see Figure~\ref{kinematics_diagram}). The spatial density
of the BLR is then given by
\begin{equation}
P(r, \theta, \phi | E, L, \theta_0) \propto \frac{1}{v} \times \frac{1}{| \sqrt{\sin^2\theta_0
-
\cos^2\theta}|},
\end{equation}
where the angular term comes from integrating over the uniform
distribution of azimuthal angle $\phi_0$ of the angular momentum
vector, and $v$ is the total magnitude of the velocity vector:
\begin{equation}
v = \sqrt{2E + \frac{2GM_{BH}}{r}}.
\end{equation}
Owing to the symmetry of our model we can consider only
$\theta_0<\pi/2$ (i.e. $L_z>0$), obtaining the following limits on the
allowed $\theta$ coordinate for the BLR:
\begin{eqnarray}
\frac{\pi}{2}-\theta_0 < \theta < \frac{\pi}{2}+\theta_0.
\end{eqnarray}
As $\theta_0$ approaches zero, the model represents a thin disk, while
as $\theta_0$ approaches $\pi/2$ the model covers the whole
sphere. Conservation of energy and angular momentum limits the radial
coordinate to the range:
\begin{equation}
r>-\frac{GM_{BH}}{2E} - \frac{1}{2} \sqrt{\left(\frac{GM_{BH}}{E}\right)^2 + \frac{2L^2}{E}},
\end{equation}
\begin{equation}
r< -\frac{GM_{BH}}{2E} + \frac{1}{2} \sqrt{\left(\frac{GM_{BH}}{E}\right)^2 + \frac{2L^2}{E}}.
\end{equation}
Finally, $E$ and $L$ are connected by the usual condition:
\begin{equation}
|L| \leq \frac{GM_{BH}}{\sqrt{-2E}}.
\end{equation}
For every allowed value of $r$, $\theta$, and $\phi$, the component of the
velocity vector along the line of sight can be computed in the standard
manner, resulting in two solutions per position, in general (outbound
and inbound; four if one considers $L_z<0$ as well). More complex
geometries and kinematics can be obtained by superpositions of
multiple sets of $E$, $L$, and $\theta_0$ values within the same potential
given by $M_{\rm BH}$. However, this further increases the
dimensionality of parameter space and computing time. Therefore in
this example we will only use one such set.

We apply prior probability distributions to the model parameters as described for the geometry model.  The priors for the extra parameters in the dynamical model not part of the geometry model are as follows.  The parameter $\theta_0$ has a flat prior in the parameter ranging from 0 to $\pi/2$.  The parameters $M_{\rm BH}$, $E$, and $L$ have flat priors in the log of the parameter.

In addition, in order to impose a BLR gas geometry, we model the
distribution of {\it illuminated gas}, as the product of the spatial
distribution given by the dynamical model with that imposed by one of
our geometrical models, representing in this case the illumination
function. This results in a broad range of geometries, giving the
model a considerable flexibility (for example, in the future one could
think of an anisotropic illumination function to model dust
obscuration). The procedure is illustrated in
Figure~\ref{dynamics_illumination}.  Note that the radial distribution
of the illuminated gas is not gaussian anymore, as in the
ring/disk/shell geometry model.  The mean radius then is not the
$r_{0}$ parameter of the geometry model, but must be computed
numerically for each set of geometric and dynamical parameters.  Similarly, the mean width is no longer $\sigma_r$ and must be computed numerically.

A model spectrum at a given time is obtained by summing all the line
of sight velocities, weighted by the spatial density of illuminated
gas multiplied by the continuum flux at an epoch corresponding to the
appropriate lag-time.  In order to compare with real data, the model
spectrum is then convolved with a gaussian to represent instrumental
broadening.  Since we do not expect real data to match our model perfectly, we introduce a relatively large uncertainty in the form of the spectral line by adding gaussian noise with a variance of $\sigma^2(F) = \alpha\,F + \beta$, where $\alpha=0.00018$ and $\beta = 0.025$.  This model for the variance assumes both a dependence on spectral line flux $F$ through the $\alpha$ parameter and a dependence on the continuum uncertainty through the $\beta$ parameter.  The units of $\alpha$ are flux and the units of $\beta$ are flux$^2$.  The specific values of $\alpha$ and $\beta$ are related to the arbitrary flux units of our simulated spectra and result in a signal to noise of $\sim 4$.  Conservatively this signal-to-noise ratio is lower than typically achieved in state of the art spectral monitoring campaigns.  Examples of synthetic spectra at a resolution of FWHM$=13.1$\,\AA, or $\sim 800$\,km\,s$^{-1}$ at the wavelength of H$\beta$, are
shown in Figure~\ref{fig_spectra}.  The face-on disk systems (top and middle panel of Figure~\ref{fig_spectra} have velocity bins of $\sim$120\,km\,s$^{-1}$ while the sphere system (bottom panel) has velocity bins of $\sim$20\,km\,s$^{-1}$. Notice how the line shapes are
clearly different even for models with the same black hole
mass. This is a clear illustration of the power of velocity resolved
reverberation mapping as a diagnostic of the BLR geometry as well as
kinematics.

\begin{figure}[h!]
\begin{center}
\includegraphics[scale=0.4]{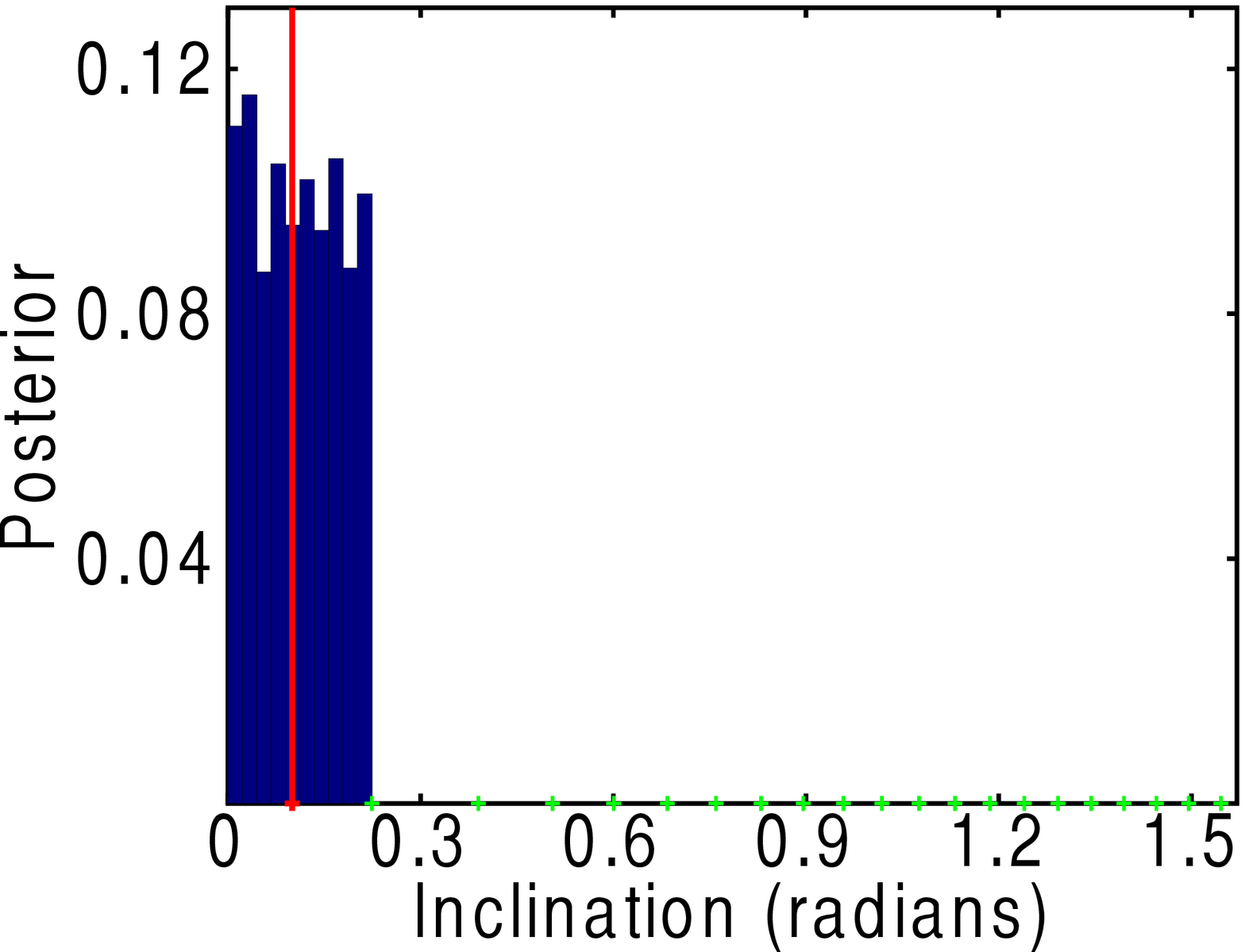}
\includegraphics[scale=0.4]{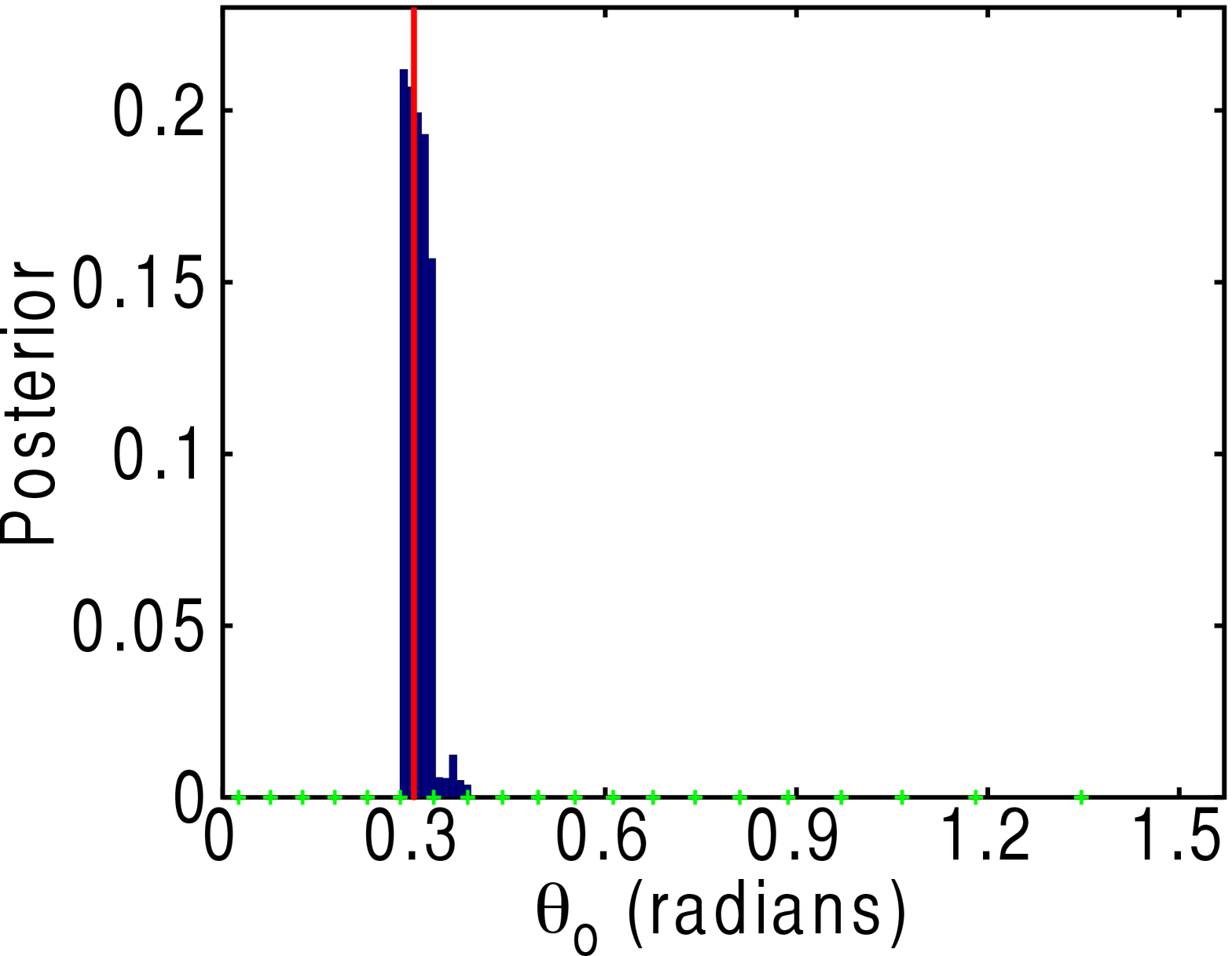}
\includegraphics[scale=0.4]{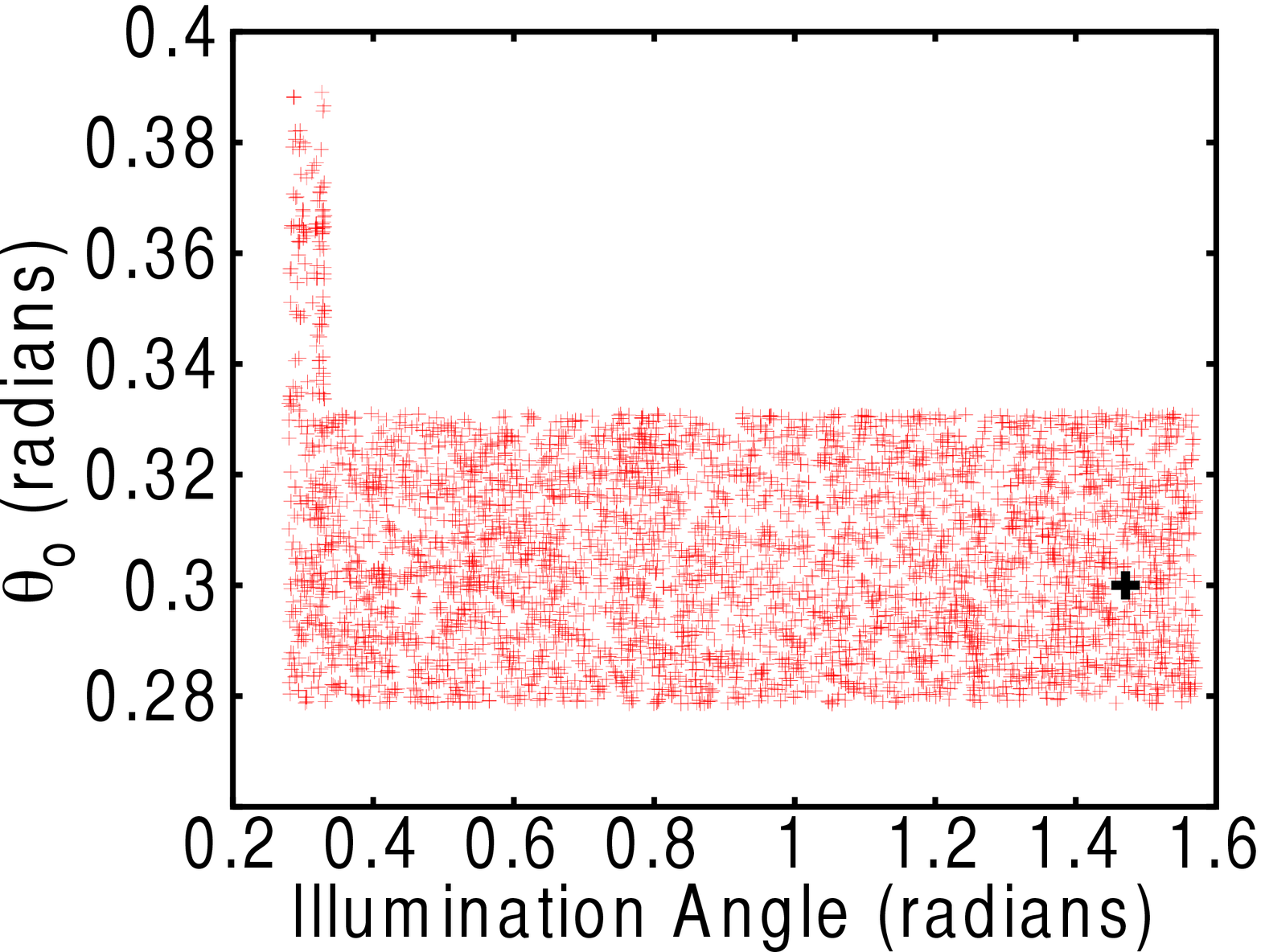}
\caption{Posterior pdfs for the first dynamical simulated dataset: face-on disk with the orbits confined to the disk.  (Top) inclination angle, (middle) $\theta_0$, and (bottom) the joint pdf of $\theta_0$ and the illumination angle.  Notice in the joint pdf that $\theta_0$ may only be larger than $\sim 0.3$\,radians when the illumination angle is $\sim 0.3$\,radians, so the angular extent of the disk is well determined.
\label{fig_dyn1b}}
\end{center}
\end{figure}

\begin{figure}[h!]
\begin{center}
\includegraphics[scale=0.4]{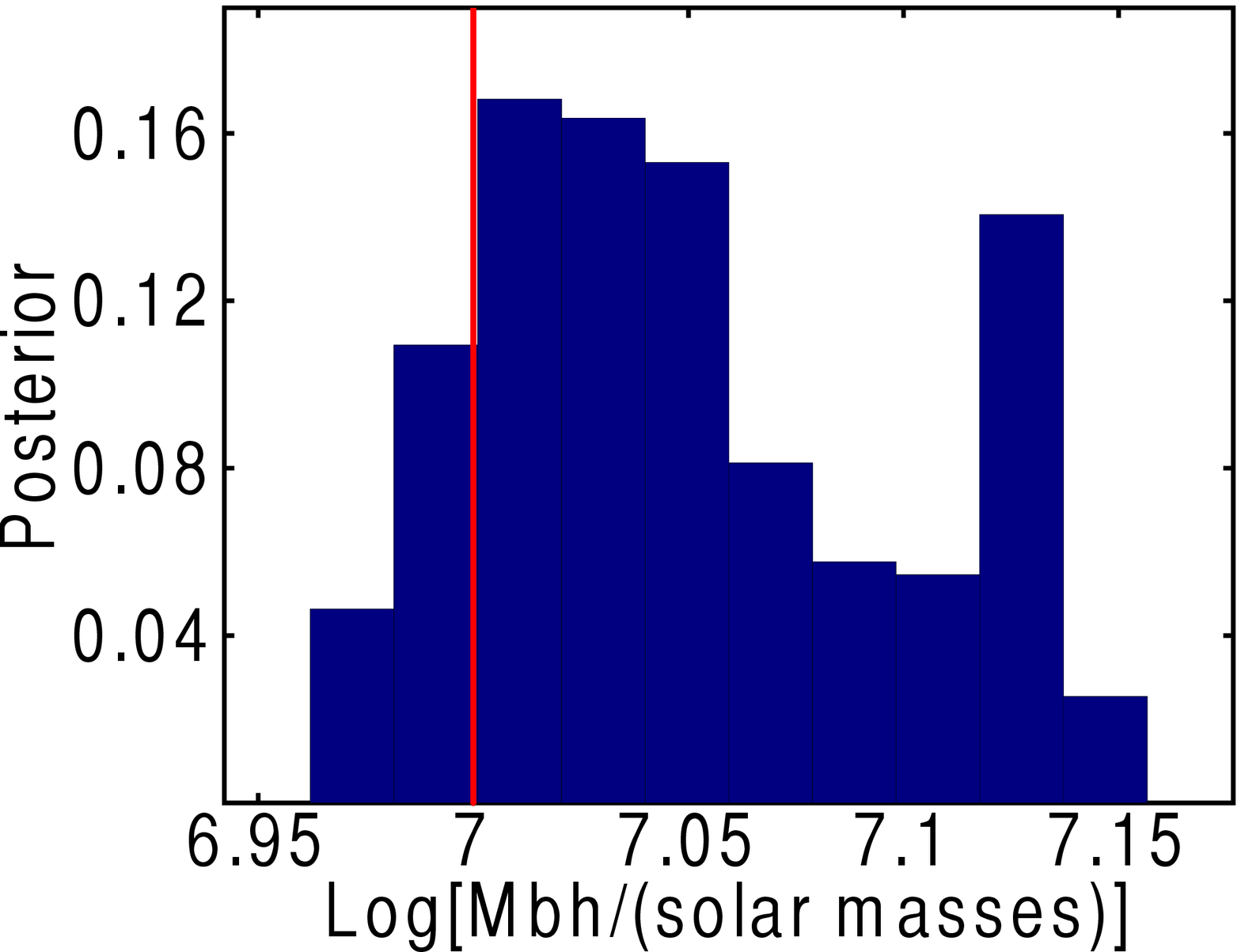}
\includegraphics[scale=0.4]{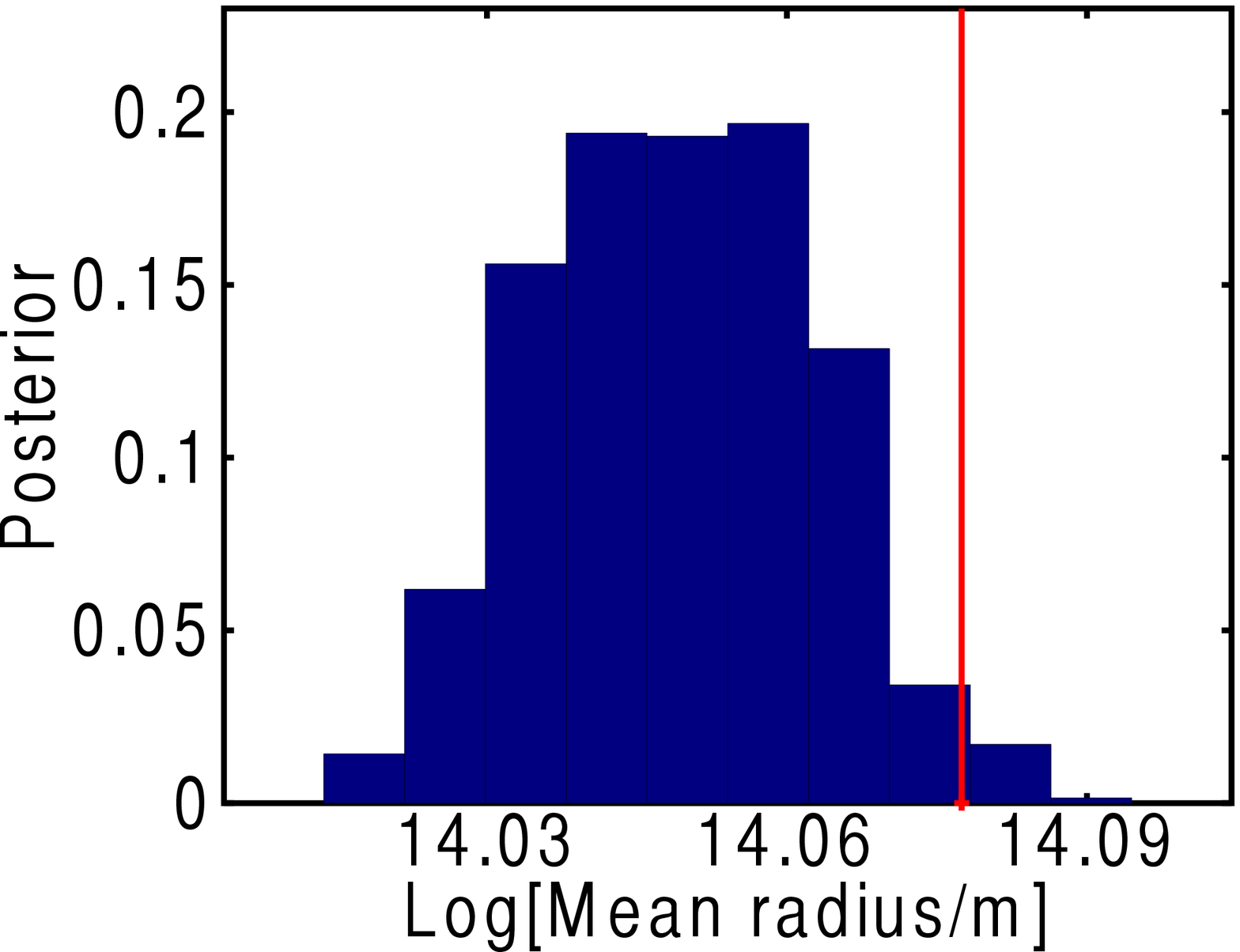}
\includegraphics[scale=0.4]{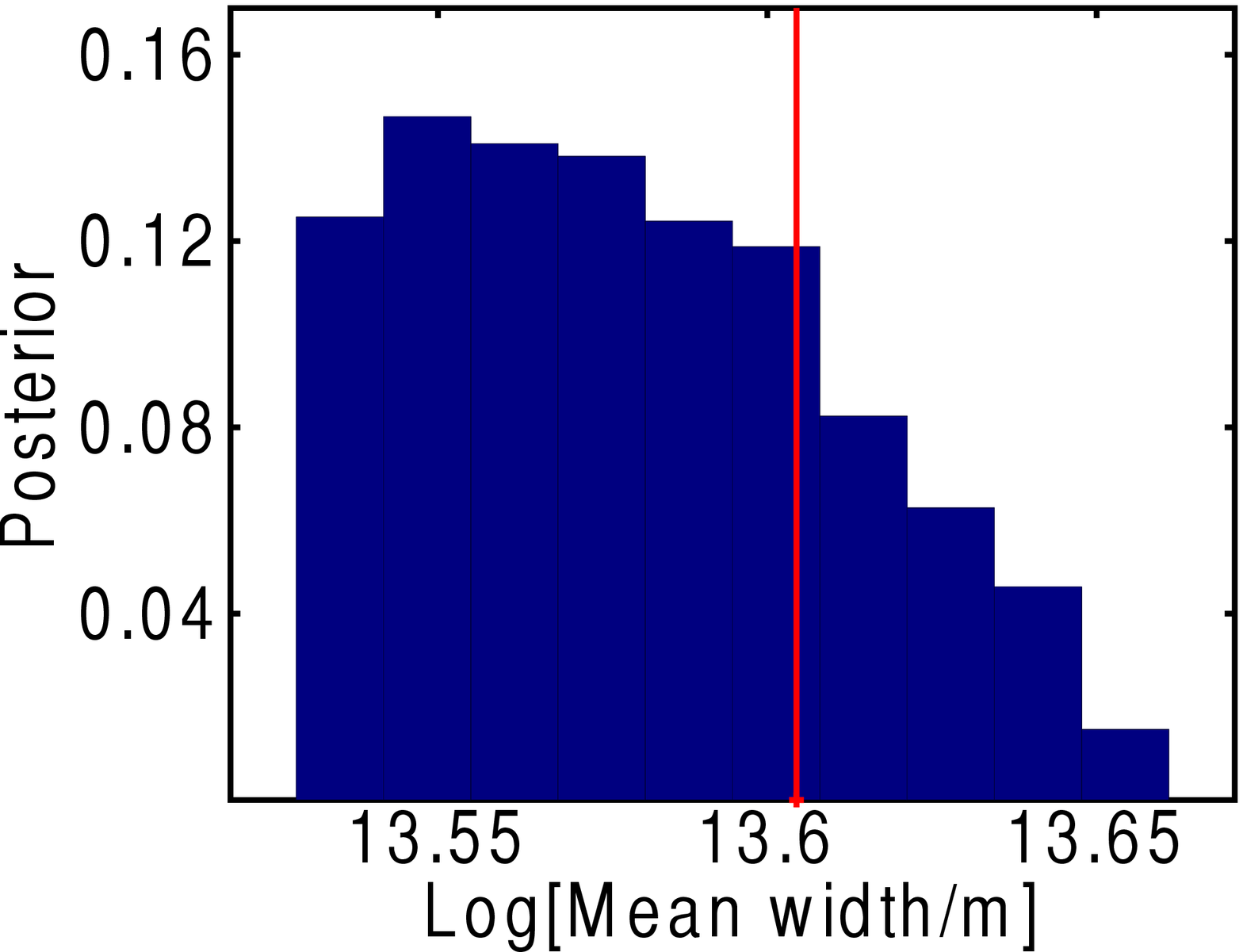}
\caption{Posterior pdfs for the second dynamical simulated dataset: face-on disk with the orbits in the entire sphere.  (Top) black hole mass, (middle) the average radius of the BLR gas mass, and (bottom) the average width of the BLR gas mass.
\label{fig_dyn2}}
\end{center}
\end{figure}

\begin{figure}[h!]
\begin{center}
\includegraphics[scale=0.4]{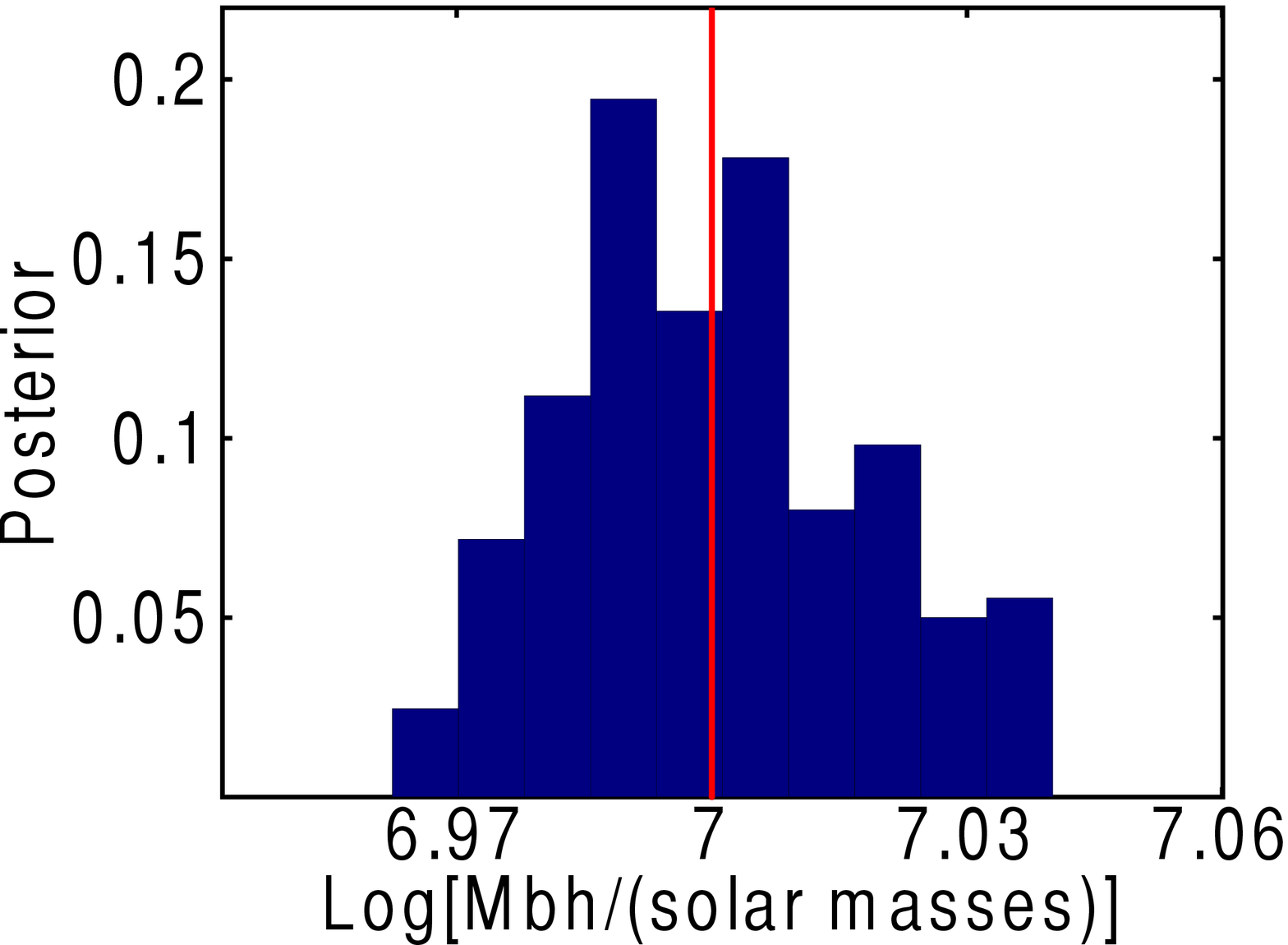}
\includegraphics[scale=0.4]{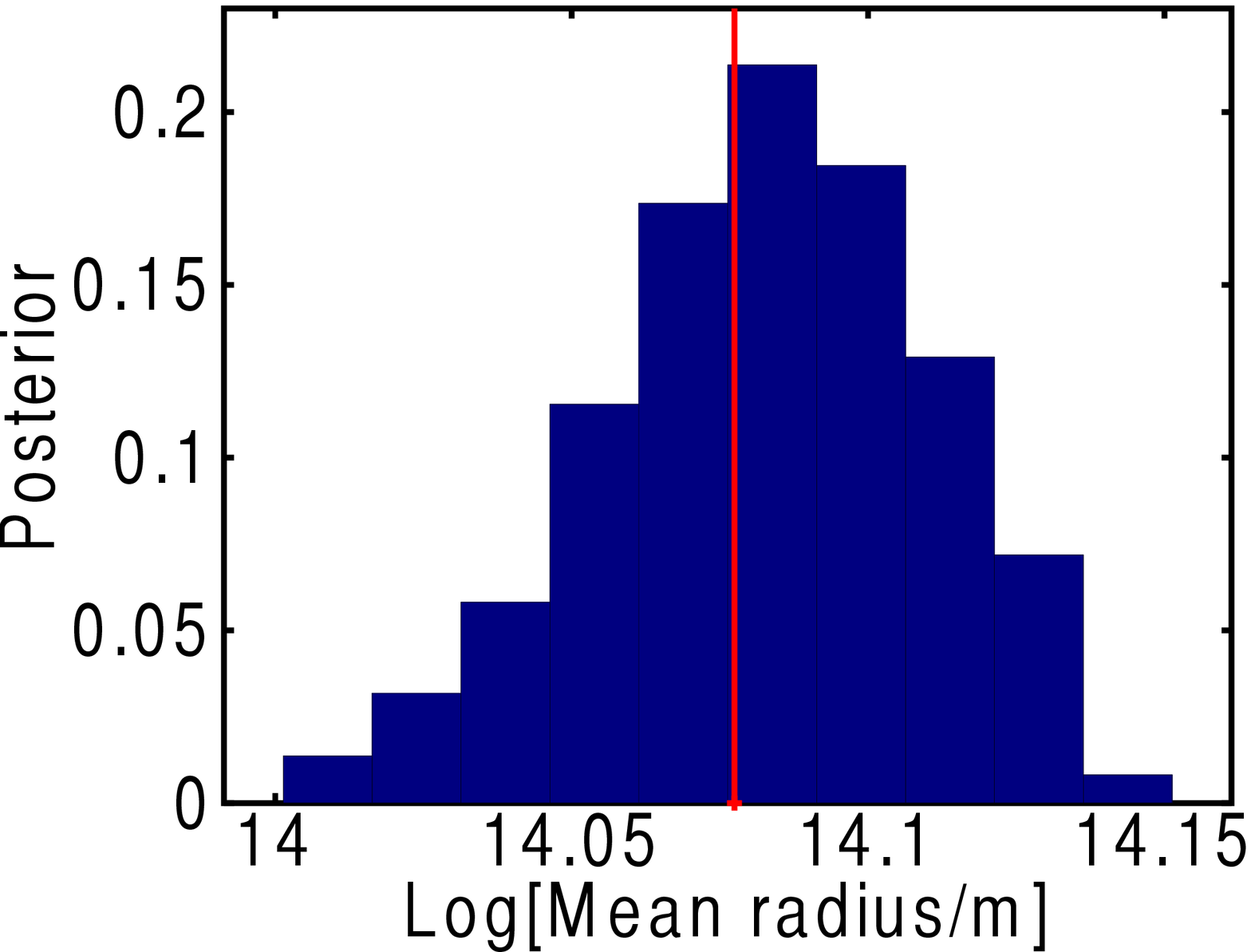}
\includegraphics[scale=0.4]{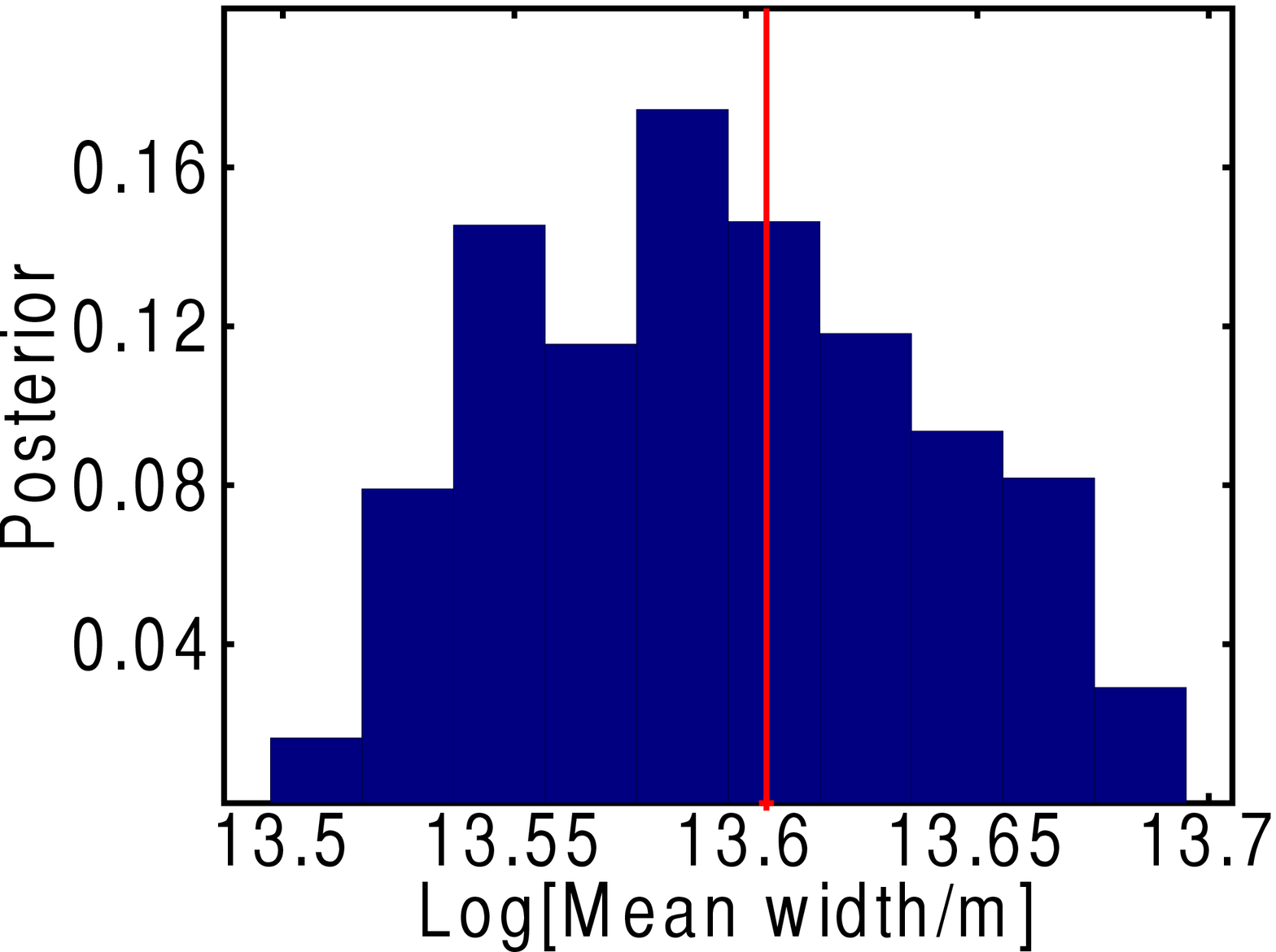}
\caption{Posterior pdfs for the third dynamical simulated dataset: sphere configuration with orbits allowed in the entire sphere.  (Top) black hole mass, (middle) the average radius of the BLR gas mass, and (bottom) the average width of the BLR gas mass.
\label{fig_dyn3}}
\end{center}
\end{figure}

\begin{figure}[h!]
\begin{center}
\includegraphics[scale=1.3]{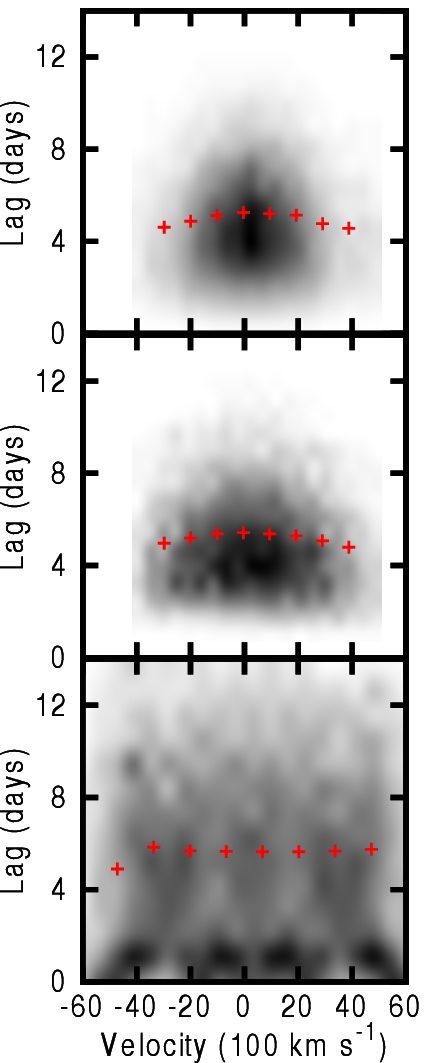}
\caption{Velocity-resolved transfer functions for the three dynamical simulated datasets: (top) face-on disk with orbits confined to the disk, (middle) face-on disk with orbits allowed in entire sphere, and (bottom) sphere configuration with orbits allowed in entire sphere.  The red crosses show the response weighted mean lag in 10 velocity bins across the spectra.
\label{fig_vres_transfunct}}
\end{center}
\end{figure}

\subsubsection{Testing the dynamical model}

We test our dynamical model by creating simulated data-sets consisting
of timeseries of the continuum flux and of the line profiles of a
broad line.  The line profiles of the simulated datasets are shown in Figure~\ref{fig_spectra}.   The kinematics parameters $E$ and $L$ are initially chosen to satisfy nearly circular orbits of
the BLR gas at the mean radius given by the illumination function.  A disk of broad line emitting material can be constrained by either the illumination function or the value of $\theta_0$.

The first simulated dataset is a thin disk viewed nearly face-on, with
dynamics imposed by a single value of energy and angular momentum.  The thin disk is constrained by the value of $\theta_0$, while the illumination function describes the whole sphere being illuminated.  This means that all allowed orbits lie in the disk and that the rest of the sphere does not contain broad line emitting gas. 
The second simulated dataset is also a thin disk viewed nearly face-on with a single value of energy and angular momentum, but for this case the illumination function constrains the disk.  We choose a value of $\theta_0$ close to $\pi/2$ so that orbits are allowed in the entire sphere.  The third simulated dataset is a fully illuminated sphere with orbits that are also allowed in the entire sphere, so again  $\theta_0$ is close to $\pi/2$.  This is still an axisymmetric configuration, as the BLR gas density imposed by the kinematics depends upon the $\theta$-coordinate.  The true parameter values of the three simulated
datasets used to test the kinematics model are shown in Table~\ref{table_simdyndata}.

We test each of the three simulated datasets assuming only one set of kinematics parameters.  The parameter values inferred using our method are shown in Table~\ref{table_simdynresults}, while the full posterior pdfs are shown for all parameters of interest for the first simulated dataset in Figures~\ref{fig_dyn1a} and \ref{fig_dyn1b}.  The posterior pdfs for the black hole mass, average radius of BLR gas mass, and average width of the BLR gas mass are also shown for the second and third simulated datasets in Figures~\ref{fig_dyn2} and \ref{fig_dyn3}.   They show that the black hole mass, average radius, and average width of the BLR are well determined for all three simulated datasets.  The angular parameters are also well determined when physically possible.  For example, for the first dynamics simulated dataset of a face-on disk with orbits confined to the disk, the inclination angle and $\theta_0$ are determined to within one or two grid points, while the illumination angle is only constrained to be $\gtrsim 0.3$\,radians.  The illumination angle cannot be determined more accurately because the BLR gas emission only comes from the disk, so as long as the entire disk is illuminated the spectrum is not sensitive to further changes in the illumination angle.

Finally, we also compute the velocity-resolved transfer functions for the three simulated datasets, shown in Figure~\ref{fig_vres_transfunct}.  As expected, the transfer functions for the face-on disk configurations show little response at very small lags, while the sphere configuration shows the highest intensity of response at small lags.  The transfer functions for the face-on disk configurations are similar, but clearly lead to different line profiles, again illustrating the power of modeling the full dataset rather than just trying to model the transfer function.

\section{Summary and Conclusions}
\label{sect_concl}
We introduce and test a new method for analyzing reverberation mapping data of AGN by directly modeling the BLR.  We illustrate our method by creating simple geometry and dynamical models of the BLR.  Using a model of the BLR geometry to reproduce the integrated line flux timeseries from reverberation mapping data allows us to estimate the average radius of the BLR, as well as the mean width, illumination function, and inclination angle to the line of sight.  Models of the BLR that include geometry {\it and} dynamical information allow us to additionally estimate the black hole mass and obtain an estimate of the extent to which the BLR gas orbits are confined to a disk or the whole sphere.  

Our method of analysis provides several advantages over previous methods.  First, previous methods rely upon cross-correlation to obtain a mean radius for the BLR and a virial relation with unknown virial coefficient to obtain an estimate of the black hole mass.  Our method estimates the black hole mass self-consistently, without the need for a virial coefficient.  Second, work modeling reverberation mapping data has previously focused on modeling the velocity-resolved or unresolved transfer function.  However the implications for the geometry and kinematics of the BLR are not clear for such analysis, as the transfer function is a function of the lag between the continuum and line emission.  Instead of modeling the transfer function and then interpreting the transfer function in terms of a geometrical or dynamical model of the BLR, we focus on modeling the BLR directly.  This allows us to extract more information and thus constrain the models more tightly.  Finally, our fast method provides estimates of the uncertainty in the model parameter values and can be used with numerical algorithms such as Nested Sampling that allow for model selection.  Our main results can be summarized as follows:

\begin{enumerate}
\item We create simulated datasets using the geometry model with known true parameter values and find that we can recover these values with uncertainties that depend upon the random uncertainty of the reverberation mapping data.  We can recover the mean radius of the BLR to within $\sim0.1$\,dex and the mean width of the BLR to within $\sim 0.2$\,dex for simulated data with an integrated line flux uncertainty of $1.5$\%.  We can also place constraints on the inclination and illumination with uncertainties of $\sim 0.2$\, radians for simulated data with face-on and spherical geometry configurations and $1.5$\% integrated line flux uncertainty.  Current integrated line flux uncertainties of about $\sim 5$\% are on the edge of what would allow for successful recovery of more than just a mean radius for the BLR.

\item We create simulated datasets using the dynamical model that consist of timeseries of a broad line profile and we compare them to mock spectra made using our model.  Despite the larger number of free parameters in our dynamical model, we find that we can recover all the parameters physically possible because the line profile is a stronger constraint on the model than the integrated line flux.  We can recover the black hole mass and the mean radius of the BLR to within $\sim0.05$\,dex, for simulated data with a line profile signal to noise ratio of $\sim4$ per spectral pixel.  We can also recover the mean width of the BLR to within $\sim0.1$\,dex and the inclination angle and illumination angle to within $\sim 2$ grid spacings over which the BLR density is defined.
\end{enumerate}

The small random uncertainties obtained in our tests of the simple
geometry and dynamical models are partly due to the inherent
assumption that our simulated data is drawn directly from the set of
possible model configurations.  In order to simulate the expected
systematic error in applying simple models to complicated real BLR
systems, we have added substantial Gaussian noise to instances of the
model in order to create our simulated datasets.  The timeseries of
line profiles, in the case of the dynamical model, is very
constraining, and leads to the reduced random uncertainty in the mean
radius and mean width of the BLR by a factor of two for the dynamical
model, as compared to the geometry model.  When applying the method to
real data we expect larger uncertainties, owing to modelling
errors. The uncertainties quoted here should therefore be considered
as lower limits to the overall precision of the method for data of
comparable quality.  This emphasizes the importance of good quality
data {\it and} increasingly more realistic models, for recovering
detailed information about the BLR from reverberation mapping data.

While we have created and tested both simple geometry and dynamical
models, our method is more general, allowing for use of any geometry
or dynamical model that can be simply parameterized.  We plan to
expand our library of models to include inflowing or outflowing BLR
gas, which may be needed to explain some of the line profile
asymmetries of current reverberation mapping data.

\acknowledgments
We thank the referee for helpful comments.  We thank our friends and collaborators in the LAMP 2008 project for many insightful conversations.  We are grateful to Chris Kochanek and Vardha Nicola Bennert for helpful suggestions on the manuscript.  We acknowledge support by the NSF through CAREER award NSF-0642621, and by the Packard Foundation through a Packard Fellowship.  AP also acknowledges support by the NSF through the Graduate Research Fellowship Program.

\clearpage

\begin{deluxetable}{ccccccc}
\tabletypesize{\scriptsize}
\tablecaption{Simulated Geometry Data True Parameter Values \label{table_simdata}}
\tablewidth{0pt}
\tablehead{ 
\colhead{Data} &
\colhead{Model} &
\colhead{Uncertainty$^{\ast}$} &
\colhead{$r_{o}$} &
\colhead{$\sigma_r$} &
\colhead{Inclination Angle} &
\colhead{Illumination Angle} \\
\colhead{} & \colhead{} & \colhead{} & \colhead{[$10^{14}$m]} & \colhead{[$10^{14}$m]} & \colhead{[radians]} & \colhead{[radians]}
}
\startdata
1  & Inclined Disk & 1.5\%  & $5$ & $0.3\,r_{\rm mean}=1.5$ & 0.79 & 0.22 \\
2  & Inclined Disk & 5\%    & $5$ & $1.5$ & 0.79 & 0.22 \\
3  & Edge-On Disk  & 1.5\%  & $5$ & $1.5$ & $\pi/2$ & 0.22 \\
4  & Face-On Disk  & 1.5\%  & $5$ & $1.5$ & 0.0 & 0.22 \\ 
5  & Shell         & 1.5\%  & $5$ & $1.5$ & -- & $\,\,\,\,\pi/2$ 
\enddata
\tablecomments{Each simulated dataset consists of 60 line emission datapoints and the same 120 continuum emission datapoints, where the line emission timeseries start half-way through the continuum timeseries.  $^{\ast}$Line flux uncertainty of the simulated dataset.}
\end{deluxetable}

\begin{deluxetable}{cccccc}
\tabletypesize{\scriptsize}
\tablecaption{Simulated Geometry Data Recovered Parameter Values \label{table_simresults}}
\tablewidth{0pt}
\tablehead{ 
\colhead{Data} &
\colhead{Model} &
\colhead{$r_{o}$} &
\colhead{$\sigma_r$} &
\colhead{Inclination Angle} &
\colhead{Illumination Angle} \\
\colhead{} & \colhead{} & \colhead{[$10^{14}$m]} & \colhead{[$10^{14}$m]} & \colhead{[radians]} & \colhead{[radians]}
}
\startdata
1  & Inclined Disk &  $4.53\pm0.47$ & $2.93\pm0.78$ & -- & -- \\
2  & Inclined Disk &  $4.81\pm1.10$ & $2.24\pm1.56$ & -- & -- \\
3  & Edge-On Disk  &  $4.76\pm0.65$ & $1.83\pm0.60$ & -- & -- \\
4  & Face-On Disk  &  $4.93\pm0.67$ & $3.10\pm1.31$ & $0.23\pm0.13$ & $0.29\pm0.10$ \\ 
5  & Shell         &  $4.48\pm0.52$ & $1.85\pm0.45$ & -- & $\,\,\,\,1.22\pm0.29$ 
\enddata
\tablecomments{Results for 150,000 iterations using an MCMC algorithm.  See Section \ref{sect_tests} for a discussion of why average values for the angular parameters are not quoted for most simulated geometry data-sets. }
\end{deluxetable}

\begin{deluxetable}{ccccccccc}
\tabletypesize{\scriptsize}
\tablecaption{Simulated Dynamics Data True Parameter Values \label{table_simdyndata}}
\tablewidth{0pt}
\tablehead{ 
\colhead{Data} &
\colhead{Model} &
\colhead{$\langle$S/N$\rangle$$^{\ast}$} &
\colhead{$M_{BH}$} &
\colhead{Mean radius} &
\colhead{Mean width} &
\colhead{Inclination Angle} &
\colhead{Illumination Angle} &
\colhead{$\theta_o$} \\
\colhead{} & \colhead{} & \colhead{} & \colhead{[$10^7 M_{\odot}$]} & \colhead{[$10^{14}$m]} & \colhead{[$10^{13}$m]} & \colhead{[radians]} & \colhead{[radians]} & \colhead{[radians]}
}
\startdata
1  & Face-on Disk &  4.6 & $1$ &  1.132 & 4.484 & 0.1 & $\pi/2$ & 0.3 \\
2  & Face-on Disk &   4.4 & $1$ &  1.195 & 4.022 & 0.1 & 0.3 & $\pi/2$ \\
3  & Sphere  &  3.5 & $1$ & 1.195 & 4.022 & 0.1 & $\pi/2$ & \,\,\,\,$\pi/2$
\enddata
\tablecomments{$^{\ast}$Average signal to noise of the line flux profile.  Each simulated dataset consists of 60 line emission profiles and the same 120 continuum emission datapoints, where the line emission timeseries start half-way through the continuum timeseries.  The simulated line emission profiles are created by taking the true model and adding gaussian noise with a variance of $v = \alpha \times Flux + \beta$, where $\alpha=0.00018$ and $\beta = 0.025$.}
\end{deluxetable}

\begin{deluxetable}{cccccccc}
\tabletypesize{\scriptsize}
\tablecaption{Simulated Dynamics Data Recovered Parameter Values \label{table_simdynresults}}
\tablewidth{0pt}
\tablehead{ 
\colhead{Data} &
\colhead{Model} &
\colhead{$M_{BH}$} &
\colhead{Mean radius} &
\colhead{Mean width} &
\colhead{Inclination Angle} &
\colhead{Illumination Angle} &
\colhead{$\theta_o$} \\
\colhead{} & \colhead{} & \colhead{[$10^7 M_{\odot}$]} & \colhead{[$10^{14}$m]} & \colhead{[$10^{13}$m]} & \colhead{[radians]} & \colhead{[radians]} & \colhead{[radians]}
}
\startdata
1  & Face-on Disk & $0.95\pm0.05$ &  $1.04\pm0.04$ & $4.27\pm0.05$ & $0.11\pm0.07$ & -- & $0.31\pm0.02$ \\
2  & Face-on Disk & $1.10\pm0.13$ &  $1.12\pm0.04$ & $3.80\pm0.03$ & $0.12\pm0.06$ & $0.31\pm0.02$ & $1.40\pm0.13$ \\
3  & Sphere           & $1.00\pm0.04$  &  $1.21\pm0.08$ & $3.95\pm0.04$ & $0.12\pm0.07$ & $1.46\pm0.06$ & \,\,\,\,$1.50\pm0.05$
\enddata
\tablecomments{Results for $470\times10^3$ (data 1), $330\times10^3$ (data 2), and $110\times10^3$ (data 3) iterations using an MCMC algorithm.  See Section \ref{sect_tests} for a discussion of why average values for the angular parameters are not quoted for the illumination angle of data 1. }
\end{deluxetable}


\begin{thebibliography}

\bibitem[\protect\citeauthoryear{Antonucci}{1993}]{antonucci93} Antonucci, R.\ 1993, \araa, 31, 473 

\bibitem[\protect\citeauthoryear{Bennert et al.}{2011}]{bennert11} Bennert, V.~N., Auger, 
M.~W., Treu, T., Woo, J.-H., \& Malkan, M.~A.\ 2011, \apj, 726, 59 

\bibitem[\protect\citeauthoryear{Bentz et al.}{2006}]{bentz06} Bentz, M.~C., Peterson, 
B.~M., Pogge, R.~W., Vestergaard, M., \& Onken, C.~A.\ 2006, \apj, 644, 133

\bibitem[\protect\citeauthoryear{Bentz et al.}{2009}]{bentz09} Bentz, M.~C., et al.\ 
2009, \apj, 705, 199

\bibitem[\protect\citeauthoryear{Bentz et al.}{2010}]{bentz10} Bentz, M.~C., et al.\ 
2010, \apjl, 720, L46 

\bibitem[\protect\citeauthoryear{Blandford \& McKee}{1982}]{blandford82} Blandford, R.~D., \& McKee, C.~F.\ 1982, \apj, 255, 419 

\bibitem[\protect\citeauthoryear{Bottorff et al.}{1997}]{bottorff97} Bottorff, M., Korista, 
K.~T., Shlosman, I., \& Blandford, R.~D.\ 1997, \apj, 479, 200 

\bibitem[\protect\citeauthoryear{Brewer, P{\'a}rtay, \& 
Cs{\'a}nyi}{2010}]{dnest} Brewer B.~J., P{\'a}rtay L.~B., Cs{\'a}nyi G., 
2010, ``Diffusive Nested Sampling'', Statistics and Computing, DOI: 10.1007/s11222-010-9198-8. arXiv:0912.2380

\bibitem[\protect\citeauthoryear{Brewer, Treu, \& 
Pancoast}{2011}]{brewer11} Brewer B.~J., et al. 2011, in prep.

\bibitem[Caticha(2008)]{caticha} Caticha, A.\,
Lectures on Probability, Entropy, and Statistical Physics, arXiv: 0808.0012

\bibitem[\protect\citeauthoryear{Collin et al.}{2006}]{collin06} Collin, S., Kawaguchi, T., Peterson, B.~M., \& Vestergaard, M.\ 2006, \aap, 456, 75 

\bibitem[\protect\citeauthoryear{Denney et al.}{2009}]{denney09} Denney, K.~D., et al.\ 
2009, \apj, 702, 1353 

\bibitem[\protect\citeauthoryear{Done \& Krolik}{1996}]{done96} Done, C., \& Krolik, J.~H.\ 1996, \apj, 463, 144

\bibitem[\protect\citeauthoryear{Elvis}{2000}]{elvis00} Elvis, M.\ 2000, \apj, 545, 63

\bibitem[\protect\citeauthoryear{Emmering et al.}{1992}]{emmering92} Emmering, R.~T., 
Blandford, R.~D., \& Shlosman, I.\ 1992, \apj, 385, 460 

\bibitem[\protect\citeauthoryear{Ferrarese \& Ford}{2005}]{ferrarese05} Ferrarese, L., \& Ford, H.\ 2005, \ssr, 116, 523 

\bibitem[\protect\citeauthoryear{Graham et al.}{2010}]{graham10} Graham, A.~W., Onken, 
C.~A., Athanassoula, E., \& Combes, F.\ 2010, arXiv:1007.3834 

\bibitem[\protect\citeauthoryear{Greene et al.}{2010}]{greene10} Greene, J.~E., Peng, 
C.~Y., \& Ludwig, R.~R.\ 2010, \apj, 709, 937 

\bibitem[\protect\citeauthoryear{Horne et al.}{2003}]{horne03} Horne, K., Korista, 
K.~T., \& Goad, M.~R.\ 2003, \mnras, 339, 367 

\bibitem[\protect\citeauthoryear{Kaspi et al.}{2000}]{kaspi00} Kaspi, S., Smith, P.~S., 
Netzer, H., Maoz, D., Jannuzi, B.~T., \& Giveon, U.\ 2000, \apj, 533, 631 

\bibitem[Kelly et al.(2009)]{2009ApJ...698..895K} Kelly, B.~C., Bechtold, 
J., \& Siemiginowska, A.\ 2009, \apj, 698, 895 

\bibitem[Koz{\l}owski et al.(2010)]{2010ApJ...708..927K} Koz{\l}owski, S., 
et al.\ 2010, \apj, 708, 927 

\bibitem[\protect\citeauthoryear{Krolik \& Done}{1995}]{krolik95} Krolik, J.~H., \& Done, C.\ 1995, \apj, 440, 166 

\bibitem[\protect\citeauthoryear{Krolik}{2001}]{krolik01} Krolik, J.~H.\ 2001, \apj, 551, 
72

\bibitem[\protect\citeauthoryear{Lynden-Bell \& Rees}{1971}]{lynden71} Lynden-Bell, D., \& Rees, M.~J.\ 1971, \mnras, 152, 461 

\bibitem[\protect\citeauthoryear{MacKay}{2003}]{2003itil.book.....M} MacKay D.~J.~C., 2003, Information Theory, Inference and Learning Algorithms. Cambridge University Press. Available online at www.inference.phy.cam.ac.uk/mackay/itila/book.html

\bibitem[\protect\citeauthoryear{Marconi et al.}{2008}]{marconi08}  Marconi, A., Axon, 
D.~J., Maiolino, R., Nagao, T., Pastorini, G., Pietrini, P., Robinson, A., 
\& Torricelli, G.\ 2008, \apj, 678, 693 

\bibitem[MacLeod et al.(2010)]{2010ApJ...721.1014M} MacLeod, C.~L., et al.\ 
2010, \apj, 721, 1014 


\bibitem[\protect\citeauthoryear{Murray et al.}{1995}]{murray95} Murray, N., Chiang, J., 
Grossman, S.~A., \& Voit, G.~M.\ 1995, \apj, 451, 498 

\bibitem[\protect\citeauthoryear{Netzer \& Marziani}{2010}]{netzer10} Netzer, H., \& Marziani, P.\ 2010, \apj, 724, 318 

\bibitem[\protect\citeauthoryear{Onken et al.}{2004}]{onken04} Onken, C.~A., Ferrarese, 
L., Merritt, D., Peterson, B.~M., Pogge, R.~W., Vestergaard, M., 
\& Wandel, A.\ 2004, \apj, 615, 645 

\bibitem[\protect\citeauthoryear{Peterson}{1993}]{peterson93} Peterson, B.~M.\ 1993, \pasp, 
105, 247 

\bibitem[\protect\citeauthoryear{Peterson et al.}{2004}]{peterson04} Peterson, B.~M., et 
al.\ 2004, \apj, 613, 682 

\bibitem[Press et al.(1992)]{1992ApJ...385..404P} Press, W.~H., Rybicki, 
G.~B., \& Hewitt, J.~N.\ 1992, \apj, 385, 404 

\bibitem[\protect\citeauthoryear{Rasmussen \& Williams}{2006}]{rasmussen} Rasmussen, C.~E. and Williams, C.~K.~I., 2006, Gaussian Processes for Machine Learning, MIT Press, Cambridge, MA, USA.

\bibitem[Rybicki 
\& Press(1992)]{1992ApJ...398..169R} Rybicki, G.~B., \& Press, W.~H.\ 1992, \apj, 398, 169 

\bibitem[Rybicki 
\& Press(1994)]{1994comp.gas..5004R} Rybicki, G.~B., \& Press, W.~H.\ 1994, Computer, 5004 

\bibitem[\protect\citeauthoryear{Sivia \& Skilling}{2006}]{sivia06} Sivia, D. S., Skilling, J. 2006. Data Analysis: A Bayesian Tutorial. 2nd Edition. Oxford University Press.

\bibitem[\protect\citeauthoryear{Urry \& Padovani}{1995}]{urry95} Urry, C.~M., \& Padovani, P.\ 1995, \pasp, 107, 803 

\bibitem[\protect\citeauthoryear{Walsh et al.}{2009}]{walsh09} Walsh, J.~L., et al.\ 
2009, \apjs, 185, 156 


\bibitem[\protect\citeauthoryear{Wandel et al.}{1999}]{wandel99} Wandel, A., Peterson, 
B.~M., \& Malkan, M.~A.\ 1999, \apj, 526, 579 

\bibitem[\protect\citeauthoryear{Webb \& Malkan}{2000}]{webb00} Webb, W., \& Malkan, M.\ 2000, \apj, 540, 652 

\bibitem[\protect\citeauthoryear{Woo et al.}{2007}]{woo07} Woo, J.-H., Treu, T., 
Malkan, M.~A., Ferry, M.~A., \& Misch, T.\ 2007, \apj, 661, 60

\bibitem[\protect\citeauthoryear{Woo et al.}{2010}]{woo10} Woo, J.-H., et al.\ 2010, \apj, 716, 269

\bibitem[\protect\citeauthoryear{Zu et al.}{2010}]{zu10} Zu, Y., Kochanek, C.~S., 
\& Peterson, B.~M.\ 2010, arXiv:1008.0641 



\end{thebibliography}
\end{document}